\documentclass[%
 reprint,
superscriptaddress,
preprintnumbers,
 amsmath,amssymb,
 aps,
]{revtex4-1}

\usepackage{graphicx}
\usepackage{hyperref}
\usepackage{multirow}
\usepackage[version=4]{mhchem}
\usepackage{float}
\usepackage{caption}
\usepackage{subcaption}
\usepackage{tikz}
\usepackage[outline]{contour}
\contourlength{0.5pt}
\definecolor{magenta}{rgb}{1,0,1}
\hypersetup{
	colorlinks   = true,
	citecolor    = blue,
	linkcolor = blue,
    urlcolor = blue
}

\tikzset{
  double arrow/.style args={#1 colored by #2 and #3}{
    -stealth,line width=#1,#2, 
    postaction={draw,-stealth,#3,line width=(#1)/3,
                shorten <=(#1)/3,shorten >=2*(#1)/3}, 
  }
}

\usepackage{braket}
\usepackage{color}

\begin{document}
\title{Piezoelectric ferromagnetism in two dimensional FeCl$_2$}
\author{Kayahan Saritas}
\affiliation{Department of Applied Physics, Yale University, New Haven, CT, 06520}
\author{Sohrab Ismail-Beigi}
\affiliation{Department of Applied Physics, Yale University, New Haven, CT, 06520}
\email{sohrab.ismail-beigi@yale.edu}

\begin{abstract}
%
We predict that monolayer FeCl$_2$ is a two-dimensional piezoelectric ferromagnet (PFM) with easy-axis magnetism and a Curie temperature of 260 K.  Our {\it ab initio} calculations combined with  data mining  reveal 2H-FeCl$_2$ as the only easy-axis 2D monolayer PFM, and that its magnetic anisotropy increases many-fold with moderate hole doping.  We develop an analysis based on magnetic anisotropies densities that explain the magnetic and doping-dependent behavior of FeCl$_2$, as well as VSe$_2$ and CrI$_3$, and can enable the design of future 2D magnetically ordered materials.
\end{abstract}

\maketitle

Controlling ferromagnetism using electric fields in a two-dimensional (2D) material offers significant scientific and technological interest \cite{Spaldin2019, Ramesh2007, Ma2011}.  Regarding magnetism, this is very challenging due to (i) typically weak magnetic exchange couplings, and (ii) the easy-axis (Ising) magnetic anisotropy (MA) required by the Mermin-Wagner theorem  is rarely achieved \cite{Mermin1966,Halperin2019}.  Nevertheless, 2D monolayers of  Cr$_2$Ge$_2$Te$_6$, CrI$_3$ and Fe$_3$GeTe$_2$ are ferromagnetic with Curie temperatures of 5 K \cite{Carteaux1995}, 45 K \cite{Huang2017b} and 126 K \cite{Roemer2020, Deng2018}, respectively, but room temperature ferromagnetism of 2D monolayers has not been achieved yet. Recent experiments report that multi-layered H-phase VSe$_2$ is an Ising (easy-axis) ferromagnet above room temperature \cite{Bonilla2018, Wang2021}, but, disappointingly, theory shows that in the monolayer limit, H-VSe$_2$ has strong ferromagnetic coupling with an easy-plane anisotropy \cite{Fuh2016b, Hennig2016}.  Open questions include the origin of easy-axis anisotropy in multilayered H-VSe$_2$ and experimental measurement of the magnetic properties of yet-to-be-synthesized monolayer H-VSe$_2$.

A 2D ferromagnetic system that is also piezoelectric will permit electric field control of magnetism via field-induced strain and structural distortions \cite{Lei2013}. One approach is to create 2D van der Waals heterostructures with separate piezoelectric and ferromagnetic layers: this approach may circumvent the weak ferromagnetism often observed in type-II multiferroics  \cite{Lei2013, Ma2011, Eerenstein2007}. However, in practice, only a fraction of the strain is transferred from the piezoelectric layer to the ferromagnetic layer in these materials \cite{Lei2013}.  Ideally, a single monolayer that combines piezoelectric and ferromagnetic properties may lead to significantly improved coupling.  

Here, we screen two-dimensional materials databases using density functional theory (DFT) to find candidate piezoelectric ferromagnetic materials (PFM) with easy-axis MA. We find only one compound, monolayer H-phase FeCl$_2$.  Excitingly, we predict that H-FeCl$_2$ has a Curie temperature ($T_c$) near 260 K and its $T_c$ and MA are tunable with strain and doping.  We  develop a theoretical analysis method for MA densities as a function of electron band energies that allows us to identify the key features of the electronic structure of Fe-dihalides, V-dichalcogenides, and Cr-trihalides that generate easy-axis or easy-plane anisotropy: e.g.,  we find that $d^6$ compounds are most likely to have easy-axis MA in H-phase 2D materials.  We also can explain the trends in MA and exchange couplings as a function of the chosen ligands.

\begin{figure}
    \centering
    \includegraphics[width=\linewidth]{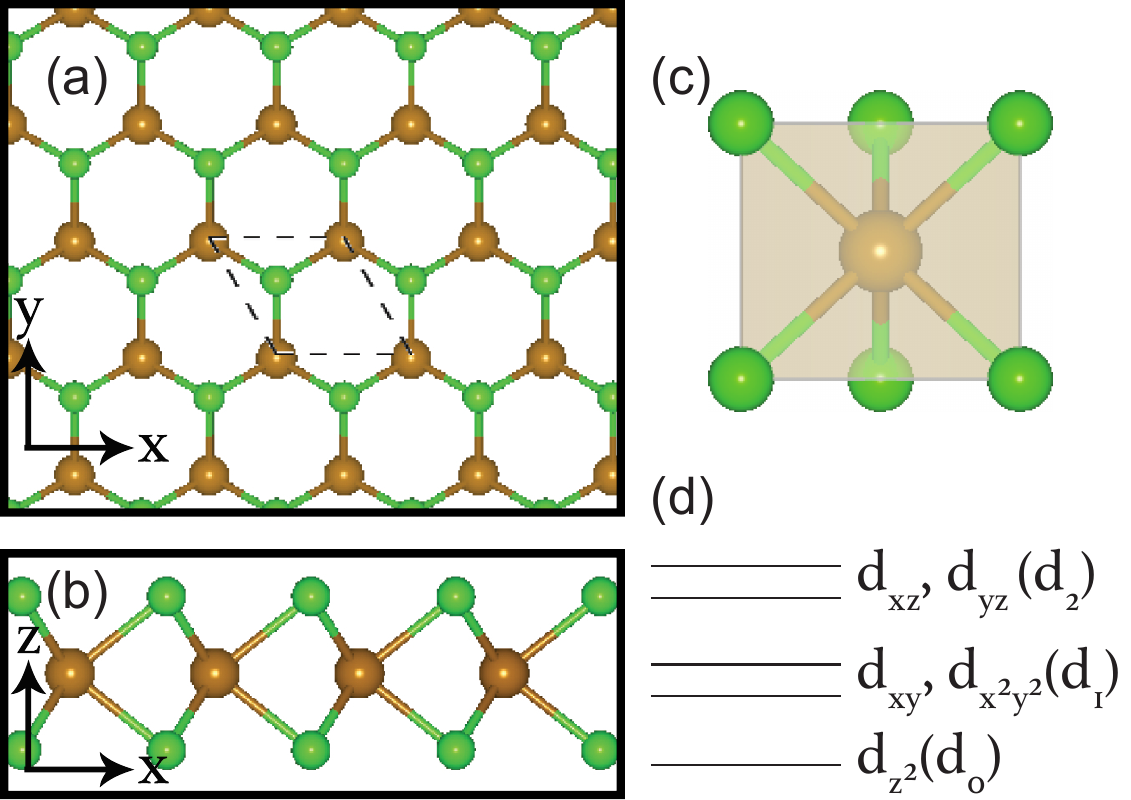}
    \caption{Atomic structure and local Fe $3d$ energy levels of 2D monolayer H-FeCl$_2$ (or H-VSe$_2$). a) Top and b) side views of the structure of 2D monolayer H-FeCl$_2$ or H-VSe$_2$.  Fe (V) are shown as gold balls and Cl (Se) are in green. The dashed line in (a) is the primitive cell. c) Trigonal prismatic arrangement of Cl (Se) atoms around the Fe (V) leads to the crystal field splittings  shown in d).}
    \label{fig:str}
\end{figure}

{\it Materials screening} ---
Our main materials discovery is that FeCl$_2$ is the only PFM as an isolated monolayer.  After screening a database of calculated magnetic ground states and electronic band gaps of more than 1600 exfoliable 2D materials \cite{Torelli}, we find that only 48 compounds can meet our screening criteria: a finite electronic bandgap, no inversion symmetry, and a finite ferromagnetic (FM) moment in the primitive cell.  Since many of the entries in the databases use very small (primitive) unit cells, the associated prediction of an FM ground state is checked using larger supercell  calculations. This leads to a further decrease in the number of screened compounds to 30. Using our magnetic stability calculations with these larger supercells, we further screen these 30 compounds by requiring a minimal energetic stability requirement of at least 10 meV/formula unit for the FM ground state relative to other magnetic orderings.   Table \ref{tab:screen} lists the resulting 14 compounds along with their key magnetic properties.  For comparison, the last three rows describe the Cr-trihalides: they lack inversion symmetry but are well-known 2D ferromagnets \cite{Kim2019a, Tiwari2021}. Only FeCl$_2$ shows positive MAE and can display Ising-type (easy axis) FM ordering in the out-of-plane direction along with a piezoelectric response.
\begin{table}[t]
    \centering
    \begin{tabular}{c|c|c|c|c}
    \hline
     & E(FM-AFM) &  MAE$^{\rm tot}$ & MAE$^{\rm tm}$ & E$_{\rm g}$ \\    
    Compound &  (meV/f.u.) &  (meV/f.u.) & (meV/f.u.) & (eV) \\
    \hline
    \hline
ScCl$_2$	&	-18	&			-0.01	&	-0.008&	0.14	\\
ScBr$_2$	&	-32	&			-0.11	&	-0.030&	0.18	\\
ScI$_2$	&	-34	&			-0.42	&	-0.060&	0.14	\\
VS$_2$	&	-48	&			-0.21	&	-0.152&	0.05	\\
VSeS	&	-52	&			-0.38	&	-0.202&	0.38	\\
VSe$_2$	&	-83	&			-0.57	&	-0.241&	0.26	\\
VTeSe	&	-69	&			-0.97	&	-0.249&	0.14	\\
VTe$_2$	&	-87	&			-1.63	&	-0.310&	0.24	\\
TiSeCl	&	-49	&			-0.21	&	-0.094&	0.09	\\
TiSeBr	&	-57	&			-0.26	&	-0.104&	0.11	\\
TiSeI	&	-49	&			-0.34	&	-0.133&	0.05	\\
FeCl$_2$	&	-182	&			{\bf0.18}	&	0.169&	0.41	\\
FeBr$_2$	&	-148	&			-0.06	&	0.397&	0.28	\\
FeI$_2$	&	-110	&			-3.99	&	0.892&	0.19	\\
\multicolumn{5}{l}{\textit{Cr-Halides}}	\\
\hline										CrCl$_3$	&	-4	&			{\bf 0.03}	&	0.015&	1.76	\\
CrBr$_3$	&	-6	&			{\bf 0.19}	&	0.032&	1.53	\\
CrI$_3$\cite{Lado2017}	&	-9	&		{\bf	0.67}	&	0.059&	1.02	\\
\hline
\hline
    \end{tabular}
    \caption{FM-AFM energies,  total magnetic anisotropy energy of the system (MAE$^{\rm tot}$), site-projected magnetic anisotropy energy on the transition metal atom only(MAE$^{\rm tm}$), electronic bandgap (E$_{\rm g}$) via SOC-DFT calculations for H-phase 2D monolayers. Positive MAE$^{\rm tot}$ values are in bold (i.e., out-of-plane or easy-axis magnetization).}
    \label{tab:screen}
\end{table}

In Table \ref{tab:screen}, we provide the energy differences between the FM state and the minimum energy antiferromagnetic state (AFM), electronic band gaps including the effect of spin-orbit coupling (SOC),  total MAE, and transition metal site projected magnetic anisotropy (MAE$^{\rm tm}$) energies (the remainder of the MAE is due to the halide or chalcogenide). Compounds in Table \ref{tab:screen}, except for Cr-halides, can be divided into two main groups depending on their electronic structure. Sc$^{2+}$, Ti$^{3+}$, and V$^{4+}$ are formally in a $d^1$ valence state, whereas Fe$^{2+}$ has a valence of $d^6$. As  these are all H-phase compounds, the trigonal prismatic crystal field surrounding the transition metal atoms leads to the ligand-field splittings of Fig. \ref{fig:str}d. Hence, the $d^1$ compounds should only have the $d_{z^2}$ level filled. We find that Fe$^{2+}$ in the Fe-halides has high-spin configuration \cite{Suppdoc}, hence in the minority spin channel only the $d_{z^2}$ state should be filled. Among the $d^1$ compounds,  FM phase stabilizes with heavier ligands, whereas  the $d^6$ compounds show the opposite trend. However, for both  $d^1$ and $d^6$, MAE decreases with heavier ligands as opposed to the increasing MAE trend in the $d^3$ Cr-halides. Finally, the relative importance of the contribution from the transition metal site MAE$^{\rm tm}$ to the total MAE$^{\rm tot}$ decreases systematically with progressively heavier ligands. 

\begin{figure}[t]
    \centering
    \includegraphics[width=\linewidth]{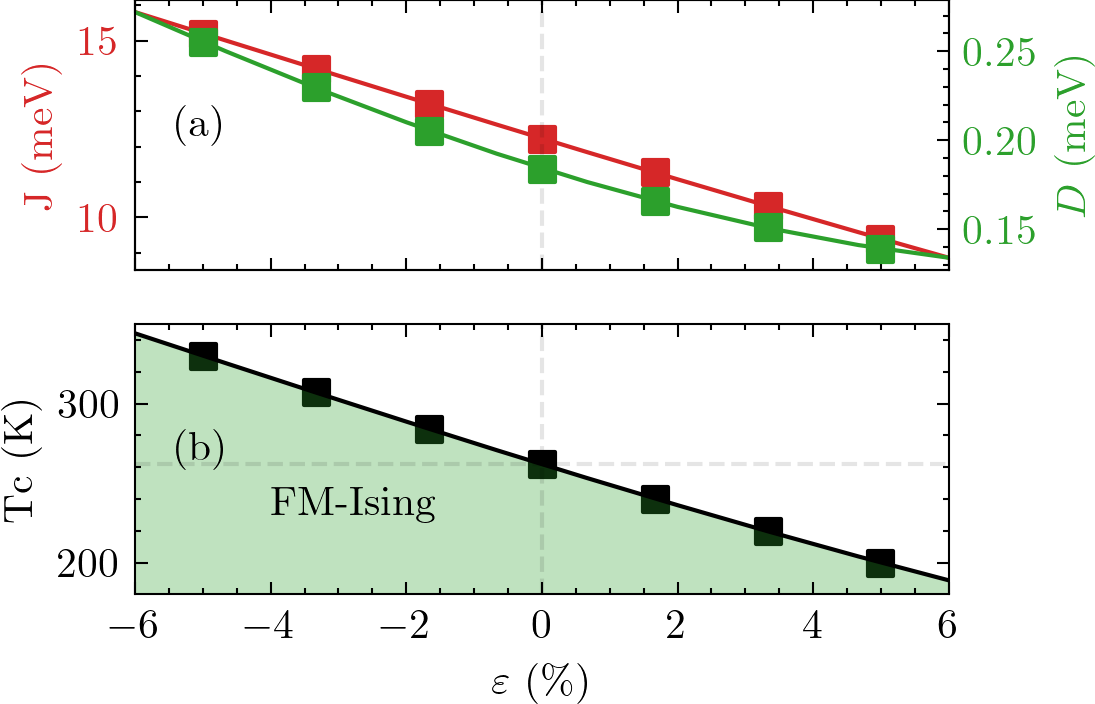}
    \caption{(a) Magnetic nearest neighbor superexchange $J$ and onsite easy-axis anisotropy $D$ of XXZ Hamiltonian (Eq. \ref{eq:ham}) computed as a function of uniaxial strain of monolayer FeCl$_2$. (b) Finite temperature phase diagram of monolayer FeCl$_2$ versus strain, $\varepsilon$, via linear spinwave theory. Compression drives FM-Ising phase, while tension drives paramagnetic phase.}
    \label{fig:tc}
\end{figure}

All compounds in Table \ref{tab:screen} (except for the Cr-trihalides) have the H-phase symmetry shown in Fig. \ref{fig:str}a-c. Most 2D materials (e.g., transition metal dichalcogenides) have two stable phases, the H and T phases. These two phases often have similar thermodynamic stabilities, and the ground state phase can depend on the environment \cite{Vettier1975, Xu2013a, Ghosh2021,Wang2021,Li2020}. 
Though H-phase FeCl$_2$ has yet to be synthesized, computational studies show that H-FeCl$_2$ is dynamically stable \cite{Zheng2018a}. Additionally, the mixed chalcogen and halide compounds in Table \ref{tab:screen},  known as Janus monolayers, exist and are under active areas of experimental and theoretical research  \cite{Lu2017,Smaili2021}.  In sum, the majority of our tabulated compounds, including H-phase FeCl$_2$, should be  experimentally realizable.

In terms of comparing Table \ref{tab:screen} to the literature, V-chalcogenides have been studied most intensively. For VS$_2$, $\Delta$=E(FM-AFM) of -47, -70 and -158 meV/f.u. were reported using PBE, LDA+U and HSE functionals, respectively \cite{Fuh2016b, Hennig2016}, while $\Delta$ of -77 and -89 meV/f.u. were found for VSe$_2$ and VTe$_2$ using PBE \cite{Fuh2016b}: all are in good agreement with Table \ref{tab:screen}.  A $\Delta$ of -249 meV was reported for FeCl$_2$ using PBE \cite{Zheng2018a}, although this is considerably larger than our value of -182 meV. Prior work has found that the magnetic coupling in VS$_2$ originates from superexchange interactions through S atoms  \cite{Hennig2016} and agrees with the Goodenough-Kanamori-Anderson rules: superexchange of orthogonal configuration is dominated by the FM contributions \cite{Goodenough1955, Kanamori1959, Blasse1965}. Investigating the densities of states in FM and AFM phases, we find that the reason FM interactions are strongly favored in Fe-halides and V-chalcogenides is due to very small crystal field splitting in these materials across the electronic gap compared to Cr-halides. Fe-halides also have stronger exchange splitting compared to V-chalcogenides which can explain increased $\Delta$ in these compounds \cite{Suppdoc}.

\begin{figure}[t]
    \centering
    \includegraphics[width=\linewidth]{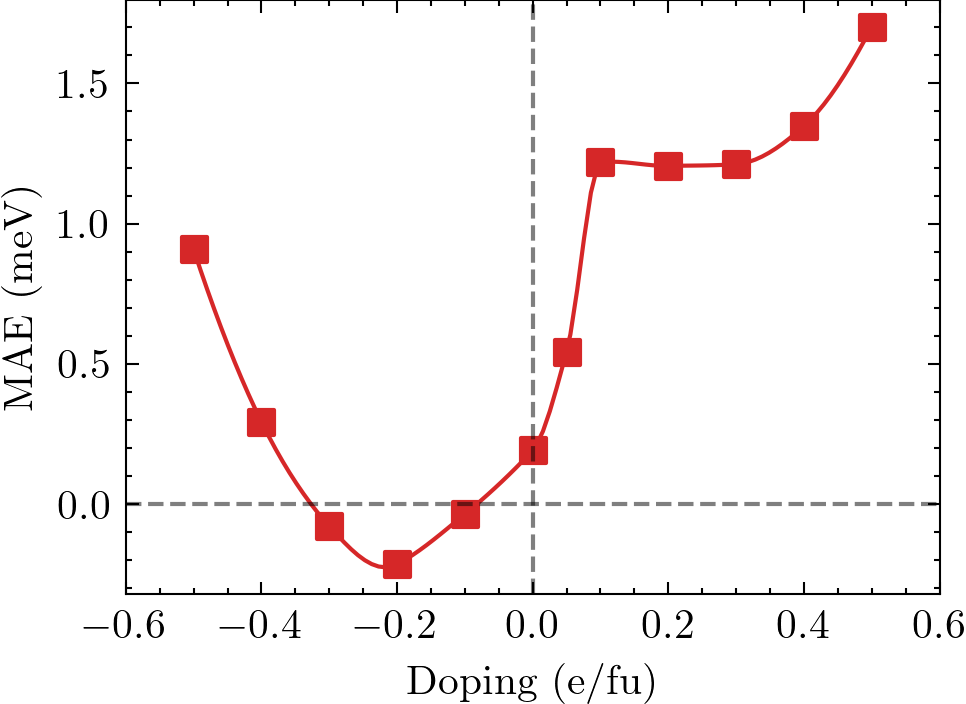}
    \caption{{\em Ab initio} MAE vs. electronic charge doping of monolayer FeCl$_2$.  Positive doping means hole doping.}
    \label{fig:doping}
\end{figure}

We find that the theoretical piezoelectric and elastic constants of FeCl$_2$ are similar to those of other 2D materials \cite{Blonsky2020, Duerloo2012}: FeCl$_2$ has a strain piezoelectric constant of $d_{11}=27.23$ pm/V, a piezoelectric constant of $e_{11}=117.5$ pC/m, and elastic coefficients $C_{11}=59.9$ and $C_{12}=18.17$ N/m. FeCl$_2$ monolayers have the $\bar{6}$m2  point group symmetry so $e_{11}$ and $d_{11}$ are the only independent piezoelectric coefficients \cite{DeJong2015}.

{\em Materials engineering} --- 
We will now discuss various strategies to manipulate and enhance the ferromagnetism of FeCl$_2$ as it is the only Ising ferromagnet in our dataset. In Fig. \ref{fig:tc}, we explain how the magnetic coupling constants and Curie temperature of monolayer FeCl$_2$ change as a function of strain. To obtain the magnetic coupling parameters, we use the XXZ spin Hamiltonian to describe the spin moments on the Fe atoms:
\begin{equation}
    \mathcal{H} = -\left[D\sum_{i}(S_i^z)^2 + \frac{J}{2}\sum_{i\neq{j}}S_iS_j + \frac{\lambda}{2}\sum_{i\neq{j}}{S_i^z}S_j^z\right]
    \label{eq:ham}\,.
\end{equation}
The sums run over Fe atoms, and pair interactions include the nearest neighbor Fe atoms. The first term describes easy-axis onsite anisotropy, and the second term is the isotropic Heisenberg exchange. When $J>0$, FM interactions are favored, whereas  $J<0$ favors AFM configurations. The last term, $\lambda$, is the anisotropic symmetric exchange that captures the magnetic anisotropy due to coupling between Fe sites via the ligands and is related to the strength of spin-orbit coupling interactions on the ligand atoms. We find that the magnetic interactions are short-ranged within the first nearest neighbors and that the anisotropic exchange is negligible in FeCl$_2$ \cite{Suppdoc}. To determine the magnetic coupling parameters from DFT, we use the FM and striped AFM phases with spins oriented in in-plane ($x$) and out-of-plane ($z$) directions. Using the $J$ and $D$ determined from the DFT calculations at each strain level, we determine the critical transition temperature ($T_c$) using linear spin-wave theory \cite{Toth2015} up to second-order terms \cite{Suppdoc} using our in-house code \cite{Suppcode}. Fig. \ref{fig:tc}(b) shows that at equilibrium the  $T_c$ is about 260 K, while at reasonable strain values, achievable using a substrate, $T_c$ can be tuned between 200 and 320 K. 

In terms of piezoelectric response, Fig. \ref{fig:tc} shows that 0.1\% strain changes $T_c$ by 1 K, and this in-plane strain can be obtained with electric fields of 3 kV/cm in MoSe$_2$ \cite{Mitra2020}. As MoSe$_2$ and FeCl$_2$ have similar $d_{11}$ coefficients, fields of a few tens of kV/cm, which are attainable in similar systems \cite{Lei2013}, will lead to significant changes in $T_c$ of order 10 K and a clear change of magnetic phase from FM to paramagnetic (PM) at a fixed temperature.  However, a full FM to PM phase change may not be necessary for applications: smaller magnetic modulations can lead to domain wall propagation in nanowires of 2D ferromagnets which can enable spintronic data storage \cite{Lei2013}.


Beyond strain, one can enhance the easy-axis MA of FeCl$_2$ via charge transfer between the monolayer and external material. Fig. \ref{fig:doping} shows the doping dependence of the MAE.  Even a modest hole doping of 0.05$e$ per formula unit makes $D$ increase threefold from 0.18 meV to 0.54 meV. A change transfer of 0.05$e$ per formula unit means a downward shift of $-0.2$ eV in the Fermi level compared to the valence band edge, therefore only the occupations of the minority spin $d$-electrons are modified.  

{\em Materials analysis and discussion} --- Although {\it ab initio} calculations of MAE (c.f. Table \ref{tab:screen}) are useful to screen materials, such calculations often only provide orbital or site decomposed MAE from integration over the entire Brillouin zone (BZ). The MAE is a delicate quantity that originates from spin-orbit interactions and depends on the distribution of valence and conduction spin and orbital states over the BZ.  Hence, an analysis of MAE requires a more fine-grained description.

A second-order perturbation theory analysis of the MAE can provide band, k-vector, and spin decomposition of the MAE for a deeper understanding of the electronic band structure \cite{Wang1993}. The second-order treatment begins with the atomic spin-orbit coupling $\propto {\bf L} \cdot {\bf S}$ on each atom ($\bf S$ and $\bf L$ are defined as spin and atomic orbital angular momentum operators). One begins with a collinear description of the spins before adding the perturbation, so the spin operator $\bf S$ acts globally on each Bloch state.  Thus the spin-orbit perturbation due to multiple atoms takes the form $H^{SOC} = - \sum_i\xi_i {\bf L}^i \cdot {\bf S}$ where $i$ runs over the atomic sites in the material with site-dependent spin-orbit constants $\xi_i$.  Physically, the spin moment is localized on the transition metal atoms: when ${\bf L}^i$ refers to the same site as $\bf S$, we have a direct SOC coupling with $\xi_i$ being the intrinsic SOC parameter for the atom; otherwise, we have an indirect SOC coupling and the associated $\xi_i$ is an effective parameter deriving from the metal-ligand hopping parameters and differences of their energy levels \cite{Kim2020a}.  Within  second-order perturbation theory, the MAE contribution from the atomic site $i$ is  \cite{Wang1993}: 
\begin{equation}
\small
    {\rm MAE}^i = {\xi_i^2}\!\!\!\sum_{vck\sigma\sigma'}\!\!\!\sigma\sigma'\frac{|\bra{vk\sigma}L^{i}_{z}\ket{ck\sigma'}|^2 - |\bra{vk\sigma}L^{i}_{x}\ket{ck\sigma'}|^2}{\epsilon_{ck\sigma'} - \epsilon_{vk\sigma}}
    \label{eq:mae}
\end{equation}
In Eq. \ref{eq:mae}, $v$ and $c$ refer to valence and conduction bands, $k$ is the reciprocal grid index, $\ket{nk\sigma}$ is a DFT Bloch state, the spin indices $\sigma,\sigma'$ take the values $\pm1$, and $\epsilon_{nk\sigma}$ are DFT band energies. To calculate the MAE of Eq. \ref{eq:mae}, we expand the valence and conduction band states of atom-centered orbitals $\ket{\phi^i_\alpha}$:  e.g., 
\[
\tiny
\bra{vk\sigma}L^{i}_{x}\ket{ck\sigma'} = \sum_{\alpha,\beta}
{\braket{vk\sigma|\phi^i_\alpha}} \bra{\phi^i_\alpha}L^{i}_{x}\ket{\phi^i_\beta}{\braket{\phi^i_\beta|ck\sigma'}}
\]
DFT postprocessing provides the inner products $\braket{\phi^i_\alpha|nk\sigma}$ for real-valued $\phi^i_\alpha$ (e.g., $d_{xz}$ or $p_y$), and we compute $\bra{\phi^i_\alpha}L^{i}_{x}\ket{\phi^i_\beta}$ by rewriting the angular character of the $\phi^i_\alpha$ as linear combinations of spherical harmonics $Y_{lm}$ and using textbook angular momentum matrices in the $Y_{lm}$ basis.
When one atom dominates the MAE, analysis of the MAE from Eq.~\ref{eq:mae} can ignore the numerical value of $\xi_i$ since it simply acts as an overall scaling.
 \begin{figure*}[t]
    \centering
    \includegraphics[width=\linewidth]{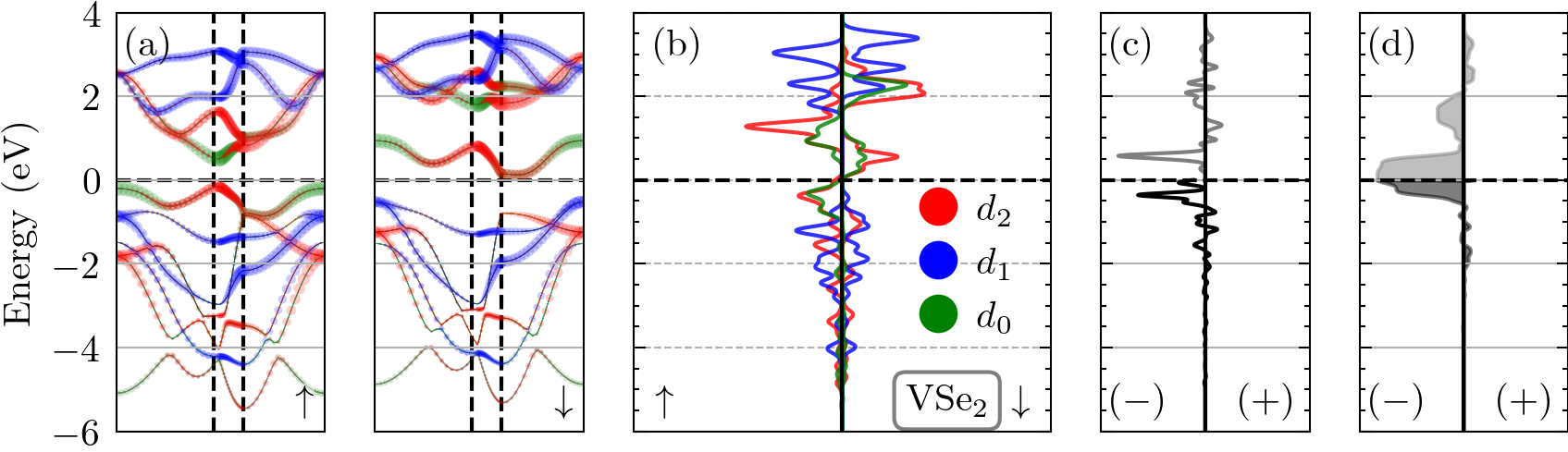}
    \includegraphics[width=\linewidth]{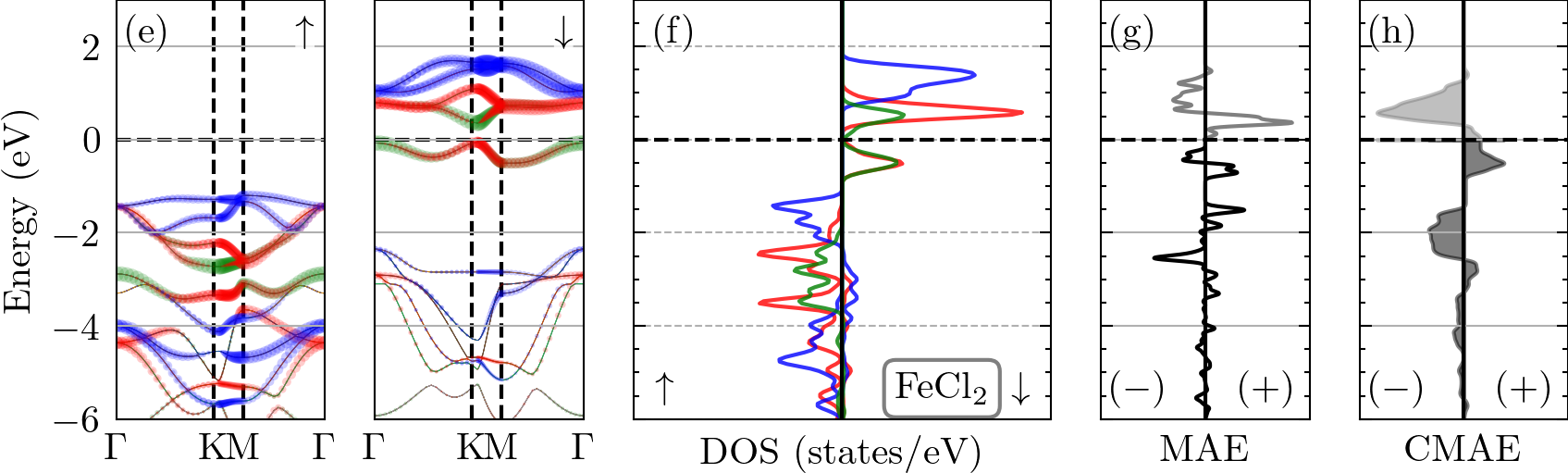}
    
    \caption{Band structures for H-phase a) VSe$_2$ and e) FeCl$_2$ projected on $d_0$=$d_{z^2}$ (green), $d_1$=$d_{xz}$/$d_{yz}$ (blue), and $d_2$=$d_{x^2-y^2}$/$d_{xy}$ (red) orbitals for majority (left panel) and minority (right panel) spins,  respectively. b) and f) show electronic densities of states (DOS) are for VSe$_2$ and FeCl$_2$, respectively. MAE densities of c)VSe$_2$  and g) FeCl$_2$, and their cumulative MAE (CMAE)  in d) and h), respectively. Valence MAE$_v(E)$ densities are given in black, while conduction MAE$_c(E)$ densities are in gray.  The Fermi energy $E_F$ is zero of energy.}
    \label{fig:band}
\end{figure*}

Eq. \ref{eq:mae} is not so useful as it stands since it produces an aggregate site-specific MAE like standard DFT calculations.  But, since it is an analytical formula as a sum over interband transitions, we can define an MAE {\it density} for pairs of valence and conduction energies to see fine-grained and energy-resolved contributions:
\begin{widetext}
\begin{equation}
\small
    {\rm MAE}^i(E_v, E_c) = {\xi_i^2}\sum_{vck\sigma \sigma'}\sigma\sigma'\frac{|\bra{vk\sigma}L^{i}_{z}\ket{ck\sigma'}|^2 - |\bra{vk\sigma}L^{i}_{x}\ket{ck\sigma'}|^2}{\epsilon_{ck\sigma'} - \epsilon_{vk\sigma}}\delta(E_v-\epsilon_{vk\sigma})\delta(E_c-\epsilon_{ck\sigma'})
    \label{eq:maevc}
\end{equation}
\end{widetext}
One can integrate this density along the valence and conduction energies to get quantities directly comparable to the density of states (DOS):  the valence  MAE density is MAE$^i_v(E_v) = \int_{E_F}^{\infty}{\rm MAE}^i(E_v, E_c)dE_c$, and the conduction MAE density is MAE$^i_c(E_c) = \int_{-\infty}^{E_F}{\rm MAE}^i(E_v, E_c)dE_v$.  One can also define cumulative integrals for each density: CMAE$^i_v(E)=\int_{-\infty}^{E}{\rm MAE}_v^i(E')dE'$ and CMAE$^i_c(E)=\int_{E}^\infty {\rm MAE}^i_c(E')dE'$.  At $E=E_F$, both are equal to the total MAE$^i$, ${\rm CMAE}_v^i(E_F)={\rm CMAE}_c^i(E_F)={\rm MAE}^i$.

We begin the analysis with a comparison and discussion of the electronic structure of VSe$_2$ and FeCl$_2$.  The trigonal prismatic symmetry of the H phase leads to the $d$-orbital crystal field levels shown in Fig.~\ref{fig:str}(d): the non-degenerate $d_{z^2}$ is lowest in  energy (denoted as $d_0$), followed by the degenerate pair $d_{xy}$/$d_{x^2-y^2}$ (denoted as $d_2$), and the degenerate pair $d_{xz}$/$d_{yz}$ ($d_1$) is highest in energy.  This ordering strictly holds at the $\Gamma$-point in the band structures of Fig.~\ref{fig:band}, but we caution in making broad arguments based on this: the orbital projections show significant mixing of different orbitals within a given band across the BZ. The lowest energy conduction and valence bands have mixed $d_0$/$d_2$ character across the BZ in  both FeCl$_2$ and VSe$_2$.  Since both monolayers are insulators, it is easy to see that Fe$^{2+}$ has  formal $d^6$ valence in FeCl$_2$ whereas V$^{4+}$ has $d^1$ valence in VSe$_2$.  A key difference between the two insulators is that the lowest energy interband transition in FeCl$_2$ is spin-preserving (minority to minority) while for VSe$_2$ it is a spin-flip process (majority to minority). Table \ref{tab:screen} shows that the Fe or V MAE completely dominates the total MAE with lighter ligands. Ligand contribution to MAE is negative, except for a small positive contribution in FeCl$_2$, which can qualitatively depend on the orthogonalization of the atomic orbitals at small magnitudes \cite{Steiner2016}. Hence, we need to focus only on the contributions from transition metal sites to MAE and respective electronic structures to understand the origin of the positive MAE in FeCl$_2$.  

Origin of the negative MAE of VSe$_2$ is relatively straightforward to understand: the valence MAE$_v$ is dominated by the highest energy valence band while the conduction MAE$_c$ is dominated by the lowest conduction band as shown in Fig. \ref{fig:band}a-d.  (The small and fluctuating valence MAE below $E_F-1$ eV in Fig. \ref{fig:band}c is due to the weak hybridization of V-$d$ and Se-$p$ states in the Se $p$-dominated valence bands.)  Both these valence/conduction bands are mixtures of $d_0$ and $d_2$ orbitals, and the only non-zero angular momentum matrix elements are among $d_2$ orbitals, $\langle d_2|L_z^{\rm V}|d_2'\rangle$ \cite{Suppdoc}.  The spin-flip nature of the low energy transition means $\sigma\sigma'=-1$ in Eq.~\ref{eq:mae} hence, giving a negative MAE$^{\rm V}$.  We argue that choices are limited to drive a positive MAE in these materials. The tensile strain would drive the V towards the atomic limit by reducing $d_0/d_2$ mixing in the low-energy valence and conduction bands and thus reduce the magnitude of the MAE.  On the other hand, enhanced exchange splitting (e.g., an enlarged $U$ acting on the $3d$ manifold) would push the minority bands higher in energy and thus turn the lowest energy transition into a spin-conserving one between majority spin bands with $\sigma\sigma'=+1$ and possibly create positive MAE.  However, typical $d^1$ ions such as Sc$^{2+}$ or Ti$^{3+}$ are expected to have smaller exchange splittings than V$^{4+}$ making this difficult to achieve, as per the examples in Table \ref{tab:screen}.  The effect of heavier ligands for strong indirect SOC will be analyzed further below.

\begin{figure}[t]
\includegraphics[width=\linewidth]{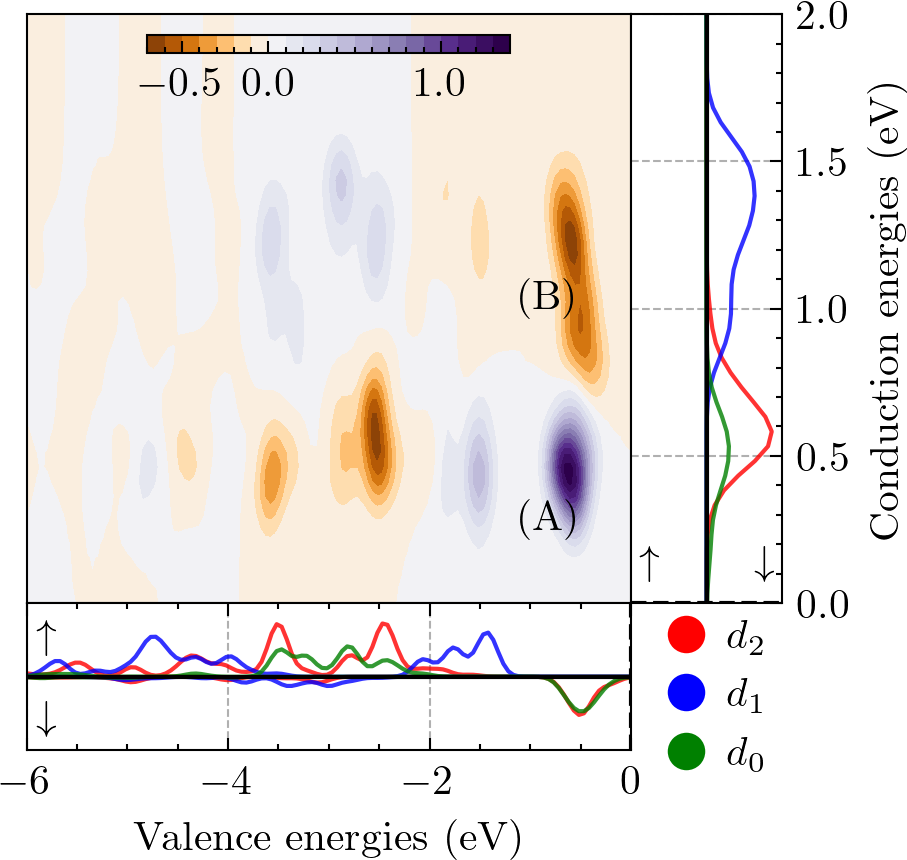}
    \caption{Two-dimensional MAE density MAE$(E_v,E_c)$ of Eq. \ref{eq:maevc} for FeCl$_2$. Purple and orange regions indicate positive and negative MAE densities, respectively, while zero density is set in white.  The peaks labeled (A) and (B) are  discussed in the text. The smaller rectangular panels overlaid on the x and y-axes show the valence (bottom) and conduction (right) electronic densities of states (DOS) which are the same as those in Fig. \ref{fig:band}.}
    \label{fig:mae2d}
\end{figure}

We now turn to the MAE analysis of the FeCl$_2$ Fe-site which is a bit more involved than VSe$_2$ V-site due to the filled majority spin Fe-$d$ states.  The valence and conduction CMAE plots in Fig. \ref{fig:band}h show that the total MAE is dominated by the highest energy valence band and the low-lying conduction bands.  The valence MAE (Fig. \ref{fig:band}g) has significant values at energies $E_F-1$ eV and lower corresponding to the low-energy filled majority spin $d$-bands, but their net integrated effect is negligible since CMAE$_v$ is very close to zero at $E_F-1$ eV. Such cancellation is expected for a full atomic shell, especially if it has full rotational symmetry; using a simple cluster model with the trigonal prismatic ligand field, we show that only a weak contribution to MAE is expected when the majority spins are fully occupied and the minority spin is empty or only the $d_0$ state is occupied \cite{Suppdoc}.  
Thus, the total MAE in FeCl$_2$ is determined by the interband transitions between the filled and empty minority bands.  
Fig. \ref{fig:band}g shows that the MAE$_c(E)$ has a prominent positive peak at $E_F+0.5$ eV and smaller but widers peak above this energy up to $E_F+1.5$ eV. The MAE$_v(E)$ has a positive peak near $E_F-0.75$ eV and a smaller negative peak around $E_F-0.25$ eV.  Hence, compared to VSe$_2$ where there was only one type of prominent interband transition, in FeCl$_2$ we have strong positive and negative MAE contributions that sum up to a net positive final answer.  To visualize this point directly, Fig. \ref{fig:mae2d} displays the MAE$(E_v,E_c)$ density (Eq. \ref{eq:maevc}) for FeCl$_2$ along with valence and conduction densities of states: a strong positive peak (A) is visible near ($E_v=-0.5$, $E_c=0.5$) coming from the interaction of the highest valence and lowest conduction band (both having $d_0+d_2$ character) while a broader negative peak (B) near ($E_v=-0.5$, $E_c=1.25$) comes from the interaction of highest valence band with higher energy conduction bands having primarily $d_1$ character.  Inspection of the $\bra{d}L_{x,z}^{\rm Fe}\ket{d'}$ matrices \cite{Suppdoc} shows that peak (A) is due to  $\bra{d_2}L_{z}^{\rm Fe}\ket{d_2'}$ while peak (B) is from $\bra{d_{0,2}}L_{x}^{\rm Fe}\ket{d_1}$ interactions. Since $\sigma\sigma'=+1$ for transitions between the same spin (minority) bands, Eq. \ref{eq:mae} then explains the signs of peaks (A) and (B).  Finally, the positive-signed interactions (A) correspond to smaller transition energies, a smaller denominator in Eq. \ref{eq:mae}, and a larger overall contribution leading to a net positive MAE.

The analysis of MAE densities also easily explains the behavior of the total MAE versus small doping seen in Fig. \ref{fig:doping}.  The MAE$_v$ in Fig. \ref{fig:band}(c) has a negative peak right below $E_F$ (peak (B) of Fig. \ref{fig:mae2d}): hole doping pushes down the Fermi energy and removes part of this negative MAE contribution.  Conversely, electron doping moves $E_F$ up in energy and removes the positive contributions of MAE$_c$ right above the Fermi energy hence driving the total MAE to smaller and eventually negative values. 

\begin{figure}[t]
    \includegraphics[width=\linewidth]{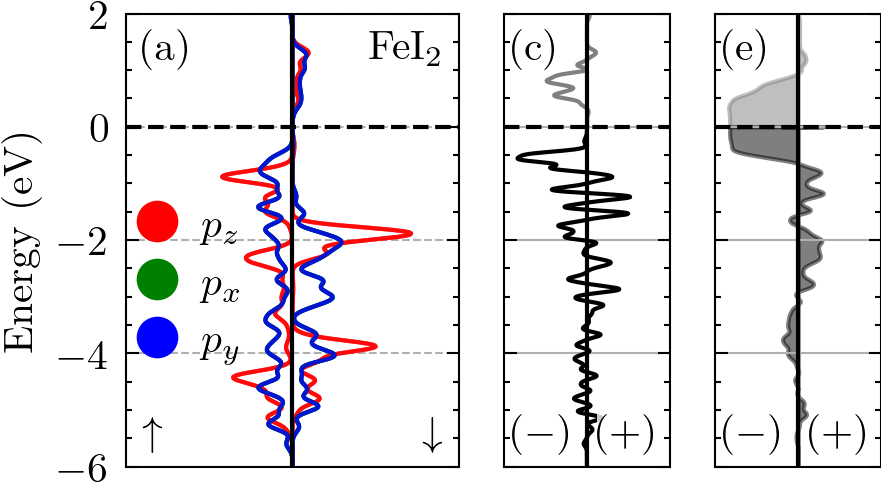}
    \includegraphics[width=\linewidth]{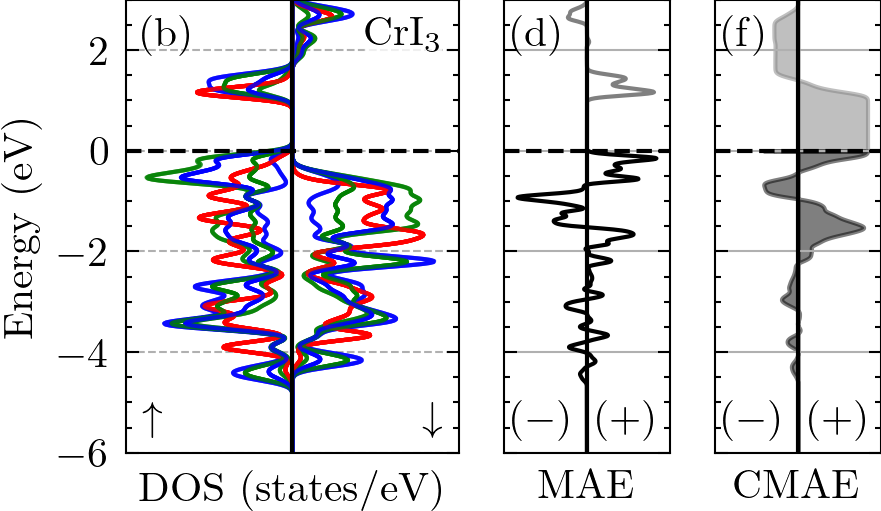}
    \caption{Projected I-5$p$ densities of states for (a) FeI$_2$ and (b) CrI$_3$. MAE densities on I sites for (c) FeI$_2$ and (d) CrI$_3$. Cumulative MAE densities on I sites for (e) FeI$_2$ and (f) CrI$_3$. Nomenclature is that of Fig. \ref{fig:band}}
    \label{fig:bandp}
\end{figure}


The MAE analysis of FeCl$_2$ and VSe$_2$ helps us extrapolate to other high-spin transition metal electronic configurations $d^n$ in trigonal prismatic symmetry.  We will assume that the primary low-energy orbital mixing is between $d_0$ and $d_2$ and that higher energy $d$ bands are mainly of $d_1$ character.  The main lesson from the $d^1$ and $d^6$ cases above is that a large exchange splitting, $\Delta_x$, would help such systems achieve positive MAE via the direct spin-orbit interaction effect on the transition metal site (not the ligands) since the lowest energy transitions would be between the same spin channels with $\sigma\sigma'=+1$ and via the $\bra{d_{2}}L_{z}\ket{d_2'}$ matrix elements. Under trigonal pyramidal symmetry, only other likely high-spin magnetic insulating phases are $d^3$, $d^5$, and $d^8$ transition metal electronic configurations. For a $d^3$ high-spin case with large $\Delta_x$, while $\sigma\sigma'=+1$ would be satisfied, the dominant interaction would now be of $\bra{d_{0,2}}L_{x}\ket{d_1}$ type leading to negative MAE. For $d^3$ with small $\Delta_x$, $\sigma\sigma'=-1$ would hold, but $\bra{d_{2}}L_{z}\ket{d_{2}}$ interactions would dominate due to their small energy denominators and give negative MAE contributions.
Stabilizing a high-spin $d^5$ configuration requires a sizeable $\Delta_x$ in the first place, and for a cluster $d^5$ model \cite{Suppdoc}, we expect the MAE to be always positive but small and scaling as  $\propto{1/\Delta_x^2}$.  The picture for $d^8$ with large $\Delta_x$ is identical to $d^3$, leading to negative MAE since the contribution from majority spin electrons would be negligible. For a  $d^8$ with small $\Delta_x$, the main low-energy contributions would come from spin-conserving $\bra{d_{0,2}}L_{x}\ket{d_1}$ and spin-flip $\bra{d_{1}}L_{z}\ket{d_1'}$  interactions, both of which give negative MAE. 

Our final analysis will explain the robust trend of decreasing MAE with heavier ligands seen in Table \ref{tab:screen}: heavy enough ligands (e.g., I or Te) which will dominate the SOC and the MAE. We can apply the perturbation analysis, Eq. \ref{eq:mae} to these systems, but we remember that we are describing an indirect SOC effect: the $\xi_i$ parameter is  modified from its pure atomic value at the ligand sites \cite{Kim2020a}, and the angular momentum operators $L^{i}_{x,z}$ act on the ligand valence $p$ orbital manifold.   

We compare CrI$_3$ and FeI$_2$ to understand their opposite signed total MAEs via an analysis of the I-5$p$ contributions. Figs. \ref{fig:bandp}a,b show spin-polarized I-5$p$ DOS above the Fermi level which stem from hybridization to the spin-polarized Fe and Cr-3$d$ states, and therefore one can get finite MAE contributions in Eq. \ref{eq:mae} from I-$p$ interband transitions.  For FeI$_2$, the low-energy conduction I-$p$ DOS is for minority spin, whereas in CrI$_3$ the low energy conduction bands are of majority spin.  The conduction MAE$_c$ densities in Figs. \ref{fig:bandp}c,d show that the total MAE is determined primarily by interactions with the lowest conduction bands ($E_F$ to about $E_F+1$ in energy).  The valence MAE$_v$ contribution is similarly controlled by interactions with the highest valence bands ($E_F-1$ to  $E_F$).  In both materials, the valence band maxima are of majority spin $p_x/p_y$ character. We have analyzed in detail the contributions from the $L_x^{\rm I}$ and $L_z^{\rm I}$ terms in Eq. \ref{eq:mae} across the BZ and find that the $L_z^{\rm I}$ matrix element contributions dominate. Since the lowest energy transition in FeI$_2$ involves a spin-flip with $\sigma\sigma'=-1$ while in CrI$_3$ it is spin conserving ($\sigma\sigma'=+1$), explaining the sign of the total MAE.  

The MAE densities in Fig. \ref{fig:bandp}(e) also show that hole doping of CrI$_3$ should lead to a decrease in the MAE (due to the removal of positive contributions near the valence band maximum) which agrees with the prior {\em ab initio} MAE results \cite{Kim2020a}. However, our analysis disagrees with their explanation of this effect: their qualitative discussion, based on band structure visualization and mean angular momenta, argues that $L_x$ interactions should dominate while we have explicitly verified that $L_z$ interactions are the key ones.  We note that MAE is a quantity derived from integration across the BZ, and with significant orbital mixing in the low-energy bands, MAE contributions in energy space are best compared to the BZ-integrated density of states.

In summary, we studied piezoelectric ferromagnetic two-dimensional monolayers using the first-principles density functional theory. Our data mining yields a single monolayer, FeCl$_2$, with a finite electronic bandgap, no inversion symmetry, and ground-state ferromagnetic ordering. Using linear spin-wave theory, we show that FeCl$_2$ has a Tc of 260 K. Magnetic anisotropy of FeCl$_2$ can be enhanced using hole doping and compressive strain.  Our analysis based on magnetic anisotropy densities directly connects the magnetic anisotropy properties of each material to key features of its electronic structure and thereby helps provide tools for the community interested in the design of  magnetic materials.

{\em Methods} --- 
All first-principles calculations for materials screening and engineering use density functional theory (DFT) with the Perdew-Burke-Ernzerhof (PBE) exchange-correlation functional \cite{pbe_perdew_generalized_1996} as implemented in the VASP 5.4.1 software \cite{Kresse1996, Kresse1996a}. We use a kinetic energy cutoff of 520 eV, periodic boundary conditions, projector-augmented wave (PAW) pseudopotentials, \cite{Kresse1999} and dipole correction \cite{Neugebauer1992}. A reciprocal grid density of 250 {\AA}$^{-3}$ is used which corresponds to 12$\times$12$\times$1 for monolayer FeCl$_2$. The MA energy (MAE) of a 2D material is
\[
{\rm MAE} = E_G([100]) - E_G([001])
\]
where $E_G(\Theta)$ is the ground state energy as a function of the spin quantization axis $\Theta$: positive MAE means  easy axis whereas negative MAE means easy-plane. The screening was automated using the atomate \cite{Mathew2017} and pymatgen \cite{Ong2013} software. 

The analysis of MAE via second-order perturbation theory used the Quantum Espresso 7.0 software \cite{Giannozzi2009}, PBE, and fully relativistic ultrasoft pseudopotentials from PSLibrary 1.0.0 \cite{DalCorso2014, Lejaeghere2016}. A plane-wave cutoff of 100 Ry and a 12$\times$12$\times$1 k-grid were found to converge total energies to 1 meV/atom.  Post-processing software tools generate the overlaps of Bloch states and real-valued atomic orbitals \cite{Suppcode}.
Curie temperatures are calculated using linear spin-wave theory \cite{Toth2015, Lado2017} using up to second-order terms \cite{Suppcode}. We calculate piezoelectric coefficients using density functional perturbation theory (DFPT) \cite{Baroni1987, Baroni2001} and elastic coefficients using finite differences. 

{\em Data availability} --- 
Structural coordinates and all the scripting/programming tools used to realize this work are available through Zenodo database with DOI number (X). The authors declare that other data related to this research are available within the paper and its Supplementary Information, or from the authors upon reasonable request.

{\em Acknowledgements} --- 
We thank the Army Research Office via grant W911NF-19-1-0371 for support of this work.   We thank the Yale Center for Research Computing for guidance and use of the research computing infrastructure.  We also thank the Extreme  Science and Engineering Discovery Environment (XSEDE), which is supported by  the National 
Science Foundation grant number ACI-1548562, for computer time on the Expanse 
supercomputer via XSEDE allocation MCA08X007.


\bibliography{main}

\begin{thebibliography}{51}%
\makeatletter
\providecommand \@ifxundefined [1]{%
 \@ifx{#1\undefined}
}%
\providecommand \@ifnum [1]{%
 \ifnum #1\expandafter \@firstoftwo
 \else \expandafter \@secondoftwo
 \fi
}%
\providecommand \@ifx [1]{%
 \ifx #1\expandafter \@firstoftwo
 \else \expandafter \@secondoftwo
 \fi
}%
\providecommand \natexlab [1]{#1}%
\providecommand \enquote  [1]{``#1''}%
\providecommand \bibnamefont  [1]{#1}%
\providecommand \bibfnamefont [1]{#1}%
\providecommand \citenamefont [1]{#1}%
\providecommand \href@noop [0]{\@secondoftwo}%
\providecommand \href [0]{\begingroup \@sanitize@url \@href}%
\providecommand \@href[1]{\@@startlink{#1}\@@href}%
\providecommand \@@href[1]{\endgroup#1\@@endlink}%
\providecommand \@sanitize@url [0]{\catcode `\\12\catcode `\$12\catcode
  `\&12\catcode `\#12\catcode `\^12\catcode `\_12\catcode `\%12\relax}%
\providecommand \@@startlink[1]{}%
\providecommand \@@endlink[0]{}%
\providecommand \url  [0]{\begingroup\@sanitize@url \@url }%
\providecommand \@url [1]{\endgroup\@href {#1}{\urlprefix }}%
\providecommand \urlprefix  [0]{URL }%
\providecommand \Eprint [0]{\href }%
\providecommand \doibase [0]{http://dx.doi.org/}%
\providecommand \selectlanguage [0]{\@gobble}%
\providecommand \bibinfo  [0]{\@secondoftwo}%
\providecommand \bibfield  [0]{\@secondoftwo}%
\providecommand \translation [1]{[#1]}%
\providecommand \BibitemOpen [0]{}%
\providecommand \bibitemStop [0]{}%
\providecommand \bibitemNoStop [0]{.\EOS\space}%
\providecommand \EOS [0]{\spacefactor3000\relax}%
\providecommand \BibitemShut  [1]{\csname bibitem#1\endcsname}%
\let\auto@bib@innerbib\@empty
\bibitem [{\citenamefont {Spaldin}\ and\ \citenamefont
  {Ramesh}(2019)}]{Spaldin2019}%
  \BibitemOpen
  \bibfield  {author} {\bibinfo {author} {\bibfnamefont {N.~A.}\ \bibnamefont
  {Spaldin}}\ and\ \bibinfo {author} {\bibfnamefont {R.}~\bibnamefont
  {Ramesh}},\ }\href {\doibase 10.1038/s41563-018-0275-2} {\bibfield  {journal}
  {\bibinfo  {journal} {Nat. Mater.}\ }\textbf {\bibinfo {volume} {18}},\
  \bibinfo {pages} {203} (\bibinfo {year} {2019})}\BibitemShut {NoStop}%
\bibitem [{\citenamefont {Ramesh}\ and\ \citenamefont
  {Spaldin}(2007)}]{Ramesh2007}%
  \BibitemOpen
  \bibfield  {author} {\bibinfo {author} {\bibfnamefont {R.}~\bibnamefont
  {Ramesh}}\ and\ \bibinfo {author} {\bibfnamefont {N.~A.}\ \bibnamefont
  {Spaldin}},\ }\href {\doibase 10.1038/nmat1805} {\bibfield  {journal}
  {\bibinfo  {journal} {Nat. Mater.}\ }\textbf {\bibinfo {volume} {6}},\
  \bibinfo {pages} {21} (\bibinfo {year} {2007})}\BibitemShut {NoStop}%
\bibitem [{\citenamefont {Ma}\ \emph {et~al.}(2011)\citenamefont {Ma},
  \citenamefont {Hu}, \citenamefont {Li},\ and\ \citenamefont {Nan}}]{Ma2011}%
  \BibitemOpen
  \bibfield  {author} {\bibinfo {author} {\bibfnamefont {J.}~\bibnamefont
  {Ma}}, \bibinfo {author} {\bibfnamefont {J.}~\bibnamefont {Hu}}, \bibinfo
  {author} {\bibfnamefont {Z.}~\bibnamefont {Li}}, \ and\ \bibinfo {author}
  {\bibfnamefont {C.-W.}\ \bibnamefont {Nan}},\ }\href {\doibase
  10.1002/adma.201003636} {\bibfield  {journal} {\bibinfo  {journal} {Adv.
  Mater.}\ }\textbf {\bibinfo {volume} {23}},\ \bibinfo {pages} {1062}
  (\bibinfo {year} {2011})}\BibitemShut {NoStop}%
\bibitem [{\citenamefont {Mermin}\ and\ \citenamefont
  {Wagner}(1966)}]{Mermin1966}%
  \BibitemOpen
  \bibfield  {author} {\bibinfo {author} {\bibfnamefont {N.~D.}\ \bibnamefont
  {Mermin}}\ and\ \bibinfo {author} {\bibfnamefont {H.}~\bibnamefont
  {Wagner}},\ }\href {\doibase 10.1103/PhysRevLett.17.1133} {\bibfield
  {journal} {\bibinfo  {journal} {Phys. Rev. Lett.}\ }\textbf {\bibinfo
  {volume} {17}},\ \bibinfo {pages} {1133} (\bibinfo {year}
  {1966})}\BibitemShut {NoStop}%
\bibitem [{\citenamefont {Halperin}(2019)}]{Halperin2019}%
  \BibitemOpen
  \bibfield  {author} {\bibinfo {author} {\bibfnamefont {B.~I.}\ \bibnamefont
  {Halperin}},\ }\href {\doibase 10.1007/s10955-018-2202-y} {\bibfield
  {journal} {\bibinfo  {journal} {J. Stat. Phys.}\ }\textbf {\bibinfo {volume}
  {175}},\ \bibinfo {pages} {521} (\bibinfo {year} {2019})}\BibitemShut
  {NoStop}%
\bibitem [{\citenamefont {Carteaux}\ \emph {et~al.}(1995)\citenamefont
  {Carteaux}, \citenamefont {Brunet}, \citenamefont {Ouvrard},\ and\
  \citenamefont {Andre}}]{Carteaux1995}%
  \BibitemOpen
  \bibfield  {author} {\bibinfo {author} {\bibfnamefont {V.}~\bibnamefont
  {Carteaux}}, \bibinfo {author} {\bibfnamefont {D.}~\bibnamefont {Brunet}},
  \bibinfo {author} {\bibfnamefont {G.}~\bibnamefont {Ouvrard}}, \ and\
  \bibinfo {author} {\bibfnamefont {G.}~\bibnamefont {Andre}},\ }\href
  {\doibase 10.1088/0953-8984/7/1/008} {\bibfield  {journal} {\bibinfo
  {journal} {J. Phys. Condens. Matter}\ }\textbf {\bibinfo {volume} {7}},\
  \bibinfo {pages} {69} (\bibinfo {year} {1995})}\BibitemShut {NoStop}%
\bibitem [{\citenamefont {Huang}\ \emph {et~al.}(2017)\citenamefont {Huang},
  \citenamefont {Clark}, \citenamefont {Navarro-Moratalla}, \citenamefont
  {Klein}, \citenamefont {Cheng}, \citenamefont {Seyler}, \citenamefont
  {Zhong}, \citenamefont {Schmidgall}, \citenamefont {McGuire}, \citenamefont
  {Cobden}, \citenamefont {Yao}, \citenamefont {Xiao}, \citenamefont
  {Jarillo-Herrero},\ and\ \citenamefont {Xu}}]{Huang2017b}%
  \BibitemOpen
  \bibfield  {author} {\bibinfo {author} {\bibfnamefont {B.}~\bibnamefont
  {Huang}}, \bibinfo {author} {\bibfnamefont {G.}~\bibnamefont {Clark}},
  \bibinfo {author} {\bibfnamefont {E.}~\bibnamefont {Navarro-Moratalla}},
  \bibinfo {author} {\bibfnamefont {D.~R.}\ \bibnamefont {Klein}}, \bibinfo
  {author} {\bibfnamefont {R.}~\bibnamefont {Cheng}}, \bibinfo {author}
  {\bibfnamefont {K.~L.}\ \bibnamefont {Seyler}}, \bibinfo {author}
  {\bibfnamefont {D.}~\bibnamefont {Zhong}}, \bibinfo {author} {\bibfnamefont
  {E.}~\bibnamefont {Schmidgall}}, \bibinfo {author} {\bibfnamefont {M.~A.}\
  \bibnamefont {McGuire}}, \bibinfo {author} {\bibfnamefont {D.~H.}\
  \bibnamefont {Cobden}}, \bibinfo {author} {\bibfnamefont {W.}~\bibnamefont
  {Yao}}, \bibinfo {author} {\bibfnamefont {D.}~\bibnamefont {Xiao}}, \bibinfo
  {author} {\bibfnamefont {P.}~\bibnamefont {Jarillo-Herrero}}, \ and\ \bibinfo
  {author} {\bibfnamefont {X.}~\bibnamefont {Xu}},\ }\href {\doibase
  10.1038/nature22391} {\bibfield  {journal} {\bibinfo  {journal} {Nature}\
  }\textbf {\bibinfo {volume} {546}},\ \bibinfo {pages} {270} (\bibinfo {year}
  {2017})}\BibitemShut {NoStop}%
\bibitem [{\citenamefont {Roemer}\ \emph {et~al.}(2020)\citenamefont {Roemer},
  \citenamefont {Liu},\ and\ \citenamefont {Zou}}]{Roemer2020}%
  \BibitemOpen
  \bibfield  {author} {\bibinfo {author} {\bibfnamefont {R.}~\bibnamefont
  {Roemer}}, \bibinfo {author} {\bibfnamefont {C.}~\bibnamefont {Liu}}, \ and\
  \bibinfo {author} {\bibfnamefont {K.}~\bibnamefont {Zou}},\ }\href {\doibase
  10.1038/s41699-020-00167-z} {\bibfield  {journal} {\bibinfo  {journal} {npj
  2D Mater. Appl.}\ }\textbf {\bibinfo {volume} {4}},\ \bibinfo {pages} {33}
  (\bibinfo {year} {2020})}\BibitemShut {NoStop}%
\bibitem [{\citenamefont {Deng}\ \emph {et~al.}(2018)\citenamefont {Deng},
  \citenamefont {Yu}, \citenamefont {Song}, \citenamefont {Zhang},
  \citenamefont {Wang}, \citenamefont {Sun}, \citenamefont {Yi}, \citenamefont
  {Wu}, \citenamefont {Wu}, \citenamefont {Zhu}, \citenamefont {Wang},
  \citenamefont {Chen},\ and\ \citenamefont {Zhang}}]{Deng2018}%
  \BibitemOpen
  \bibfield  {author} {\bibinfo {author} {\bibfnamefont {Y.}~\bibnamefont
  {Deng}}, \bibinfo {author} {\bibfnamefont {Y.}~\bibnamefont {Yu}}, \bibinfo
  {author} {\bibfnamefont {Y.}~\bibnamefont {Song}}, \bibinfo {author}
  {\bibfnamefont {J.}~\bibnamefont {Zhang}}, \bibinfo {author} {\bibfnamefont
  {N.~Z.}\ \bibnamefont {Wang}}, \bibinfo {author} {\bibfnamefont
  {Z.}~\bibnamefont {Sun}}, \bibinfo {author} {\bibfnamefont {Y.}~\bibnamefont
  {Yi}}, \bibinfo {author} {\bibfnamefont {Y.~Z.}\ \bibnamefont {Wu}}, \bibinfo
  {author} {\bibfnamefont {S.}~\bibnamefont {Wu}}, \bibinfo {author}
  {\bibfnamefont {J.}~\bibnamefont {Zhu}}, \bibinfo {author} {\bibfnamefont
  {J.}~\bibnamefont {Wang}}, \bibinfo {author} {\bibfnamefont {X.~H.}\
  \bibnamefont {Chen}}, \ and\ \bibinfo {author} {\bibfnamefont
  {Y.}~\bibnamefont {Zhang}},\ }\href {\doibase 10.1038/s41586-018-0626-9}
  {\bibfield  {journal} {\bibinfo  {journal} {Nature}\ }\textbf {\bibinfo
  {volume} {563}},\ \bibinfo {pages} {94} (\bibinfo {year} {2018})},\ \Eprint
  {http://arxiv.org/abs/1803.02038} {arXiv:1803.02038} \BibitemShut {NoStop}%
\bibitem [{\citenamefont {Bonilla}\ \emph {et~al.}(2018)\citenamefont
  {Bonilla}, \citenamefont {Kolekar}, \citenamefont {Ma}, \citenamefont {Diaz},
  \citenamefont {Kalappattil}, \citenamefont {Das}, \citenamefont {Eggers},
  \citenamefont {Gutierrez}, \citenamefont {Phan},\ and\ \citenamefont
  {Batzill}}]{Bonilla2018}%
  \BibitemOpen
  \bibfield  {author} {\bibinfo {author} {\bibfnamefont {M.}~\bibnamefont
  {Bonilla}}, \bibinfo {author} {\bibfnamefont {S.}~\bibnamefont {Kolekar}},
  \bibinfo {author} {\bibfnamefont {Y.}~\bibnamefont {Ma}}, \bibinfo {author}
  {\bibfnamefont {H.~C.}\ \bibnamefont {Diaz}}, \bibinfo {author}
  {\bibfnamefont {V.}~\bibnamefont {Kalappattil}}, \bibinfo {author}
  {\bibfnamefont {R.}~\bibnamefont {Das}}, \bibinfo {author} {\bibfnamefont
  {T.}~\bibnamefont {Eggers}}, \bibinfo {author} {\bibfnamefont {H.~R.}\
  \bibnamefont {Gutierrez}}, \bibinfo {author} {\bibfnamefont {M.~H.}\
  \bibnamefont {Phan}}, \ and\ \bibinfo {author} {\bibfnamefont
  {M.}~\bibnamefont {Batzill}},\ }\href {\doibase 10.1038/s41565-018-0063-9}
  {\bibfield  {journal} {\bibinfo  {journal} {Nat. Nanotechnol.}\ }\textbf
  {\bibinfo {volume} {13}},\ \bibinfo {pages} {289} (\bibinfo {year}
  {2018})}\BibitemShut {NoStop}%
\bibitem [{\citenamefont {Wang}\ \emph {et~al.}(2021)\citenamefont {Wang},
  \citenamefont {Li}, \citenamefont {Li}, \citenamefont {Wu}, \citenamefont
  {Che}, \citenamefont {Chen},\ and\ \citenamefont {Cui}}]{Wang2021}%
  \BibitemOpen
  \bibfield  {author} {\bibinfo {author} {\bibfnamefont {X.}~\bibnamefont
  {Wang}}, \bibinfo {author} {\bibfnamefont {D.}~\bibnamefont {Li}}, \bibinfo
  {author} {\bibfnamefont {Z.}~\bibnamefont {Li}}, \bibinfo {author}
  {\bibfnamefont {C.}~\bibnamefont {Wu}}, \bibinfo {author} {\bibfnamefont
  {C.-M.}\ \bibnamefont {Che}}, \bibinfo {author} {\bibfnamefont
  {G.}~\bibnamefont {Chen}}, \ and\ \bibinfo {author} {\bibfnamefont
  {X.}~\bibnamefont {Cui}},\ }\href {\doibase 10.1021/acsnano.1c05232}
  {\bibfield  {journal} {\bibinfo  {journal} {ACS Nano}\ }\textbf {\bibinfo
  {volume} {15}},\ \bibinfo {pages} {16236} (\bibinfo {year}
  {2021})}\BibitemShut {NoStop}%
\bibitem [{\citenamefont {Fuh}\ \emph {et~al.}(2016)\citenamefont {Fuh},
  \citenamefont {Chang}, \citenamefont {Wang}, \citenamefont {Evans},
  \citenamefont {Chantrell},\ and\ \citenamefont {Jeng}}]{Fuh2016b}%
  \BibitemOpen
  \bibfield  {author} {\bibinfo {author} {\bibfnamefont {H.~R.}\ \bibnamefont
  {Fuh}}, \bibinfo {author} {\bibfnamefont {C.~R.}\ \bibnamefont {Chang}},
  \bibinfo {author} {\bibfnamefont {Y.~K.}\ \bibnamefont {Wang}}, \bibinfo
  {author} {\bibfnamefont {R.~F.}\ \bibnamefont {Evans}}, \bibinfo {author}
  {\bibfnamefont {R.~W.}\ \bibnamefont {Chantrell}}, \ and\ \bibinfo {author}
  {\bibfnamefont {H.~T.}\ \bibnamefont {Jeng}},\ }\href {\doibase
  10.1038/srep32625} {\bibfield  {journal} {\bibinfo  {journal} {Sci. Rep.}\
  }\textbf {\bibinfo {volume} {6}},\ \bibinfo {pages} {1} (\bibinfo {year}
  {2016})}\BibitemShut {NoStop}%
\bibitem [{\citenamefont {Zhuang}\ and\ \citenamefont
  {Hennig}(2016)}]{Hennig2016}%
  \BibitemOpen
  \bibfield  {author} {\bibinfo {author} {\bibfnamefont {H.~L.}\ \bibnamefont
  {Zhuang}}\ and\ \bibinfo {author} {\bibfnamefont {R.~G.}\ \bibnamefont
  {Hennig}},\ }\href {\doibase 10.1103/PhysRevB.93.054429} {\bibfield
  {journal} {\bibinfo  {journal} {Phys. Rev. B}\ }\textbf {\bibinfo {volume}
  {93}},\ \bibinfo {pages} {054429} (\bibinfo {year} {2016})}\BibitemShut
  {NoStop}%
\bibitem [{\citenamefont {Lei}\ \emph {et~al.}(2013)\citenamefont {Lei},
  \citenamefont {Devolder}, \citenamefont {Agnus}, \citenamefont {Aubert},
  \citenamefont {Daniel}, \citenamefont {Kim}, \citenamefont {Zhao},
  \citenamefont {Trypiniotis}, \citenamefont {Cowburn}, \citenamefont
  {Chappert}, \citenamefont {Ravelosona},\ and\ \citenamefont
  {Lecoeur}}]{Lei2013}%
  \BibitemOpen
  \bibfield  {author} {\bibinfo {author} {\bibfnamefont {N.}~\bibnamefont
  {Lei}}, \bibinfo {author} {\bibfnamefont {T.}~\bibnamefont {Devolder}},
  \bibinfo {author} {\bibfnamefont {G.}~\bibnamefont {Agnus}}, \bibinfo
  {author} {\bibfnamefont {P.}~\bibnamefont {Aubert}}, \bibinfo {author}
  {\bibfnamefont {L.}~\bibnamefont {Daniel}}, \bibinfo {author} {\bibfnamefont
  {J.~V.}\ \bibnamefont {Kim}}, \bibinfo {author} {\bibfnamefont
  {W.}~\bibnamefont {Zhao}}, \bibinfo {author} {\bibfnamefont {T.}~\bibnamefont
  {Trypiniotis}}, \bibinfo {author} {\bibfnamefont {R.~P.}\ \bibnamefont
  {Cowburn}}, \bibinfo {author} {\bibfnamefont {C.}~\bibnamefont {Chappert}},
  \bibinfo {author} {\bibfnamefont {D.}~\bibnamefont {Ravelosona}}, \ and\
  \bibinfo {author} {\bibfnamefont {P.}~\bibnamefont {Lecoeur}},\ }\href
  {\doibase 10.1038/ncomms2386} {\bibfield  {journal} {\bibinfo  {journal}
  {Nat. Commun.}\ }\textbf {\bibinfo {volume} {4}},\ \bibinfo {pages} {1}
  (\bibinfo {year} {2013})}\BibitemShut {NoStop}%
\bibitem [{\citenamefont {Eerenstein}\ \emph {et~al.}(2007)\citenamefont
  {Eerenstein}, \citenamefont {Wiora}, \citenamefont {Prieto}, \citenamefont
  {Scott},\ and\ \citenamefont {Mathur}}]{Eerenstein2007}%
  \BibitemOpen
  \bibfield  {author} {\bibinfo {author} {\bibfnamefont {W.}~\bibnamefont
  {Eerenstein}}, \bibinfo {author} {\bibfnamefont {M.}~\bibnamefont {Wiora}},
  \bibinfo {author} {\bibfnamefont {J.~L.}\ \bibnamefont {Prieto}}, \bibinfo
  {author} {\bibfnamefont {J.~F.}\ \bibnamefont {Scott}}, \ and\ \bibinfo
  {author} {\bibfnamefont {N.~D.}\ \bibnamefont {Mathur}},\ }\href {\doibase
  10.1038/nmat1886} {\bibfield  {journal} {\bibinfo  {journal} {Nat. Mater.}\
  }\textbf {\bibinfo {volume} {6}},\ \bibinfo {pages} {348} (\bibinfo {year}
  {2007})}\BibitemShut {NoStop}%
\bibitem [{\citenamefont {Torelli}\ \emph {et~al.}(2020)\citenamefont
  {Torelli}, \citenamefont {Moustafa}, \citenamefont {Jacobsen},\ and\
  \citenamefont {Olsen}}]{Torelli}%
  \BibitemOpen
  \bibfield  {author} {\bibinfo {author} {\bibfnamefont {D.}~\bibnamefont
  {Torelli}}, \bibinfo {author} {\bibfnamefont {H.}~\bibnamefont {Moustafa}},
  \bibinfo {author} {\bibfnamefont {K.~W.}\ \bibnamefont {Jacobsen}}, \ and\
  \bibinfo {author} {\bibfnamefont {T.}~\bibnamefont {Olsen}},\ }\href
  {\doibase 10.1038/s41524-020-00428-x} {\bibfield  {journal} {\bibinfo
  {journal} {npj Comput. Mater.}\ }\textbf {\bibinfo {volume} {6}},\ \bibinfo
  {pages} {158} (\bibinfo {year} {2020})}\BibitemShut {NoStop}%
\bibitem [{\citenamefont {Kim}\ \emph {et~al.}(2019)\citenamefont {Kim},
  \citenamefont {Yang}, \citenamefont {Li}, \citenamefont {Jiang},
  \citenamefont {Jin}, \citenamefont {Tao}, \citenamefont {Nichols},
  \citenamefont {Sfigakis}, \citenamefont {Zhong}, \citenamefont {Li},
  \citenamefont {Tian}, \citenamefont {Cory}, \citenamefont {Miao},
  \citenamefont {Shan}, \citenamefont {Mak}, \citenamefont {Lei}, \citenamefont
  {Sun}, \citenamefont {Zhao},\ and\ \citenamefont {Tsen}}]{Kim2019a}%
  \BibitemOpen
  \bibfield  {author} {\bibinfo {author} {\bibfnamefont {H.~H.}\ \bibnamefont
  {Kim}}, \bibinfo {author} {\bibfnamefont {B.}~\bibnamefont {Yang}}, \bibinfo
  {author} {\bibfnamefont {S.}~\bibnamefont {Li}}, \bibinfo {author}
  {\bibfnamefont {S.}~\bibnamefont {Jiang}}, \bibinfo {author} {\bibfnamefont
  {C.}~\bibnamefont {Jin}}, \bibinfo {author} {\bibfnamefont {Z.}~\bibnamefont
  {Tao}}, \bibinfo {author} {\bibfnamefont {G.}~\bibnamefont {Nichols}},
  \bibinfo {author} {\bibfnamefont {F.}~\bibnamefont {Sfigakis}}, \bibinfo
  {author} {\bibfnamefont {S.}~\bibnamefont {Zhong}}, \bibinfo {author}
  {\bibfnamefont {C.}~\bibnamefont {Li}}, \bibinfo {author} {\bibfnamefont
  {S.}~\bibnamefont {Tian}}, \bibinfo {author} {\bibfnamefont {D.~G.}\
  \bibnamefont {Cory}}, \bibinfo {author} {\bibfnamefont {G.~X.}\ \bibnamefont
  {Miao}}, \bibinfo {author} {\bibfnamefont {J.}~\bibnamefont {Shan}}, \bibinfo
  {author} {\bibfnamefont {K.~F.}\ \bibnamefont {Mak}}, \bibinfo {author}
  {\bibfnamefont {H.}~\bibnamefont {Lei}}, \bibinfo {author} {\bibfnamefont
  {K.}~\bibnamefont {Sun}}, \bibinfo {author} {\bibfnamefont {L.}~\bibnamefont
  {Zhao}}, \ and\ \bibinfo {author} {\bibfnamefont {A.~W.}\ \bibnamefont
  {Tsen}},\ }\href {\doibase 10.1073/pnas.1902100116} {\bibfield  {journal}
  {\bibinfo  {journal} {Proc. Natl. Acad. Sci. U. S. A.}\ }\textbf {\bibinfo
  {volume} {166}},\ \bibinfo {pages} {11131} (\bibinfo {year}
  {2019})}\BibitemShut {NoStop}%
\bibitem [{\citenamefont {Tiwari}\ \emph {et~al.}(2021)\citenamefont {Tiwari},
  \citenamefont {{Van De Put}}, \citenamefont {Sor{\'{e}}e},\ and\
  \citenamefont {Vandenberghe}}]{Tiwari2021}%
  \BibitemOpen
  \bibfield  {author} {\bibinfo {author} {\bibfnamefont {S.}~\bibnamefont
  {Tiwari}}, \bibinfo {author} {\bibfnamefont {M.~L.}\ \bibnamefont {{Van De
  Put}}}, \bibinfo {author} {\bibfnamefont {B.}~\bibnamefont {Sor{\'{e}}e}}, \
  and\ \bibinfo {author} {\bibfnamefont {W.~G.}\ \bibnamefont {Vandenberghe}},\
  }\href {\doibase 10.1103/PhysRevB.103.014432} {\bibfield  {journal} {\bibinfo
   {journal} {Phys. Rev. B}\ }\textbf {\bibinfo {volume} {103}},\ \bibinfo
  {pages} {14432} (\bibinfo {year} {2021})}\BibitemShut {NoStop}%
\bibitem [{\citenamefont {Lado}\ and\ \citenamefont
  {Fern{\'{a}}ndez-Rossier}(2017)}]{Lado2017}%
  \BibitemOpen
  \bibfield  {author} {\bibinfo {author} {\bibfnamefont {J.~L.}\ \bibnamefont
  {Lado}}\ and\ \bibinfo {author} {\bibfnamefont {J.}~\bibnamefont
  {Fern{\'{a}}ndez-Rossier}},\ }\href {\doibase 10.1088/2053-1583/aa75ed}
  {\bibfield  {journal} {\bibinfo  {journal} {2D Materials}\ }\textbf {\bibinfo
  {volume} {4}},\ \bibinfo {pages} {035002} (\bibinfo {year}
  {2017})}\BibitemShut {NoStop}%
\bibitem [{Sup({\natexlab{a}})}]{Suppdoc}%
  \BibitemOpen
  \href@noop {} {} ({\natexlab{a}}),\ \bibinfo {note} {link to SI}\BibitemShut
  {NoStop}%
\bibitem [{\citenamefont {Vettier}\ and\ \citenamefont
  {Yelon}(1975)}]{Vettier1975}%
  \BibitemOpen
  \bibfield  {author} {\bibinfo {author} {\bibfnamefont {C.}~\bibnamefont
  {Vettier}}\ and\ \bibinfo {author} {\bibfnamefont {W.~B.}\ \bibnamefont
  {Yelon}},\ }\href {\doibase 10.1016/0022-3697(75)90065-7} {\bibfield
  {journal} {\bibinfo  {journal} {J. Phys. Chem. Solids}\ }\textbf {\bibinfo
  {volume} {36}},\ \bibinfo {pages} {401} (\bibinfo {year} {1975})}\BibitemShut
  {NoStop}%
\bibitem [{\citenamefont {Xu}\ \emph {et~al.}(2013)\citenamefont {Xu},
  \citenamefont {Chen}, \citenamefont {Li}, \citenamefont {Wu}, \citenamefont
  {Guo}, \citenamefont {Zhao}, \citenamefont {Wu},\ and\ \citenamefont
  {Xie}}]{Xu2013a}%
  \BibitemOpen
  \bibfield  {author} {\bibinfo {author} {\bibfnamefont {K.}~\bibnamefont
  {Xu}}, \bibinfo {author} {\bibfnamefont {P.}~\bibnamefont {Chen}}, \bibinfo
  {author} {\bibfnamefont {X.}~\bibnamefont {Li}}, \bibinfo {author}
  {\bibfnamefont {C.}~\bibnamefont {Wu}}, \bibinfo {author} {\bibfnamefont
  {Y.}~\bibnamefont {Guo}}, \bibinfo {author} {\bibfnamefont {J.}~\bibnamefont
  {Zhao}}, \bibinfo {author} {\bibfnamefont {X.}~\bibnamefont {Wu}}, \ and\
  \bibinfo {author} {\bibfnamefont {Y.}~\bibnamefont {Xie}},\ }\href {\doibase
  10.1002/anie.201304337} {\bibfield  {journal} {\bibinfo  {journal} {Angew.
  Chemie - Int. Ed.}\ }\textbf {\bibinfo {volume} {52}},\ \bibinfo {pages}
  {10477} (\bibinfo {year} {2013})}\BibitemShut {NoStop}%
\bibitem [{\citenamefont {Ghosh}\ \emph {et~al.}(2021)\citenamefont {Ghosh},
  \citenamefont {Jose},\ and\ \citenamefont {Kumari}}]{Ghosh2021}%
  \BibitemOpen
  \bibfield  {author} {\bibinfo {author} {\bibfnamefont {R.~K.}\ \bibnamefont
  {Ghosh}}, \bibinfo {author} {\bibfnamefont {A.}~\bibnamefont {Jose}}, \ and\
  \bibinfo {author} {\bibfnamefont {G.}~\bibnamefont {Kumari}},\ }\href
  {\doibase 10.1103/PHYSREVB.103.054409} {\bibfield  {journal} {\bibinfo
  {journal} {Phys. Rev. B}\ }\textbf {\bibinfo {volume} {103}},\ \bibinfo
  {pages} {054409} (\bibinfo {year} {2021})}\BibitemShut {NoStop}%
\bibitem [{\citenamefont {Li}\ \emph {et~al.}(2020)\citenamefont {Li},
  \citenamefont {Wang}, \citenamefont {Kan}, \citenamefont {He}, \citenamefont
  {Li}, \citenamefont {Hao}, \citenamefont {Zhao}, \citenamefont {Wu},
  \citenamefont {Jin},\ and\ \citenamefont {Cui}}]{Li2020}%
  \BibitemOpen
  \bibfield  {author} {\bibinfo {author} {\bibfnamefont {D.}~\bibnamefont
  {Li}}, \bibinfo {author} {\bibfnamefont {X.}~\bibnamefont {Wang}}, \bibinfo
  {author} {\bibfnamefont {C.~M.}\ \bibnamefont {Kan}}, \bibinfo {author}
  {\bibfnamefont {D.}~\bibnamefont {He}}, \bibinfo {author} {\bibfnamefont
  {Z.}~\bibnamefont {Li}}, \bibinfo {author} {\bibfnamefont {Q.}~\bibnamefont
  {Hao}}, \bibinfo {author} {\bibfnamefont {H.}~\bibnamefont {Zhao}}, \bibinfo
  {author} {\bibfnamefont {C.}~\bibnamefont {Wu}}, \bibinfo {author}
  {\bibfnamefont {C.}~\bibnamefont {Jin}}, \ and\ \bibinfo {author}
  {\bibfnamefont {X.}~\bibnamefont {Cui}},\ }\href {\doibase
  10.1021/acsami.0c04449} {\bibfield  {journal} {\bibinfo  {journal} {ACS Appl.
  Mater. Interfaces}\ }\textbf {\bibinfo {volume} {12}},\ \bibinfo {pages}
  {25143} (\bibinfo {year} {2020})}\BibitemShut {NoStop}%
\bibitem [{\citenamefont {Zheng}\ \emph {et~al.}(2018)\citenamefont {Zheng},
  \citenamefont {Han}, \citenamefont {Zheng},\ and\ \citenamefont
  {Yan}}]{Zheng2018a}%
  \BibitemOpen
  \bibfield  {author} {\bibinfo {author} {\bibfnamefont {H.}~\bibnamefont
  {Zheng}}, \bibinfo {author} {\bibfnamefont {H.}~\bibnamefont {Han}}, \bibinfo
  {author} {\bibfnamefont {J.}~\bibnamefont {Zheng}}, \ and\ \bibinfo {author}
  {\bibfnamefont {Y.}~\bibnamefont {Yan}},\ }\href {\doibase
  10.1016/j.ssc.2017.12.025} {\bibfield  {journal} {\bibinfo  {journal} {Solid
  State Commun.}\ }\textbf {\bibinfo {volume} {271}},\ \bibinfo {pages} {66}
  (\bibinfo {year} {2018})}\BibitemShut {NoStop}%
\bibitem [{\citenamefont {Lu}\ \emph {et~al.}(2017)\citenamefont {Lu},
  \citenamefont {Zhu}, \citenamefont {Xiao}, \citenamefont {Chuu},
  \citenamefont {Han}, \citenamefont {Chiu}, \citenamefont {Cheng},
  \citenamefont {Yang}, \citenamefont {Wei}, \citenamefont {Yang},
  \citenamefont {Wang}, \citenamefont {Sokaras}, \citenamefont {Nordlund},
  \citenamefont {Yang}, \citenamefont {Muller}, \citenamefont {Chou},
  \citenamefont {Zhang},\ and\ \citenamefont {Li}}]{Lu2017}%
  \BibitemOpen
  \bibfield  {author} {\bibinfo {author} {\bibfnamefont {A.~Y.}\ \bibnamefont
  {Lu}}, \bibinfo {author} {\bibfnamefont {H.}~\bibnamefont {Zhu}}, \bibinfo
  {author} {\bibfnamefont {J.}~\bibnamefont {Xiao}}, \bibinfo {author}
  {\bibfnamefont {C.~P.}\ \bibnamefont {Chuu}}, \bibinfo {author}
  {\bibfnamefont {Y.}~\bibnamefont {Han}}, \bibinfo {author} {\bibfnamefont
  {M.~H.}\ \bibnamefont {Chiu}}, \bibinfo {author} {\bibfnamefont {C.~C.}\
  \bibnamefont {Cheng}}, \bibinfo {author} {\bibfnamefont {C.~W.}\ \bibnamefont
  {Yang}}, \bibinfo {author} {\bibfnamefont {K.~H.}\ \bibnamefont {Wei}},
  \bibinfo {author} {\bibfnamefont {Y.}~\bibnamefont {Yang}}, \bibinfo {author}
  {\bibfnamefont {Y.}~\bibnamefont {Wang}}, \bibinfo {author} {\bibfnamefont
  {D.}~\bibnamefont {Sokaras}}, \bibinfo {author} {\bibfnamefont
  {D.}~\bibnamefont {Nordlund}}, \bibinfo {author} {\bibfnamefont
  {P.}~\bibnamefont {Yang}}, \bibinfo {author} {\bibfnamefont {D.~A.}\
  \bibnamefont {Muller}}, \bibinfo {author} {\bibfnamefont {M.~Y.}\
  \bibnamefont {Chou}}, \bibinfo {author} {\bibfnamefont {X.}~\bibnamefont
  {Zhang}}, \ and\ \bibinfo {author} {\bibfnamefont {L.~J.}\ \bibnamefont
  {Li}},\ }\href {\doibase 10.1038/nnano.2017.100} {\bibfield  {journal}
  {\bibinfo  {journal} {Nat. Nanotechnol.}\ }\textbf {\bibinfo {volume} {12}},\
  \bibinfo {pages} {744} (\bibinfo {year} {2017})}\BibitemShut {NoStop}%
\bibitem [{\citenamefont {Smaili}\ \emph {et~al.}(2021)\citenamefont {Smaili},
  \citenamefont {Laref}, \citenamefont {Garcia}, \citenamefont
  {Schwingenschl\"ogl}, \citenamefont {Roche},\ and\ \citenamefont
  {Manchon}}]{Smaili2021}%
  \BibitemOpen
  \bibfield  {author} {\bibinfo {author} {\bibfnamefont {I.}~\bibnamefont
  {Smaili}}, \bibinfo {author} {\bibfnamefont {S.}~\bibnamefont {Laref}},
  \bibinfo {author} {\bibfnamefont {J.~H.}\ \bibnamefont {Garcia}}, \bibinfo
  {author} {\bibfnamefont {U.}~\bibnamefont {Schwingenschl\"ogl}}, \bibinfo
  {author} {\bibfnamefont {S.}~\bibnamefont {Roche}}, \ and\ \bibinfo {author}
  {\bibfnamefont {A.}~\bibnamefont {Manchon}},\ }\href {\doibase
  10.1103/PhysRevB.104.104415} {\bibfield  {journal} {\bibinfo  {journal}
  {Phys. Rev. B}\ }\textbf {\bibinfo {volume} {104}},\ \bibinfo {pages}
  {104415} (\bibinfo {year} {2021})}\BibitemShut {NoStop}%
\bibitem [{\citenamefont {Goodenough}(1955)}]{Goodenough1955}%
  \BibitemOpen
  \bibfield  {author} {\bibinfo {author} {\bibfnamefont {J.~B.}\ \bibnamefont
  {Goodenough}},\ }\href {\doibase 10.1103/PhysRev.100.564} {\bibfield
  {journal} {\bibinfo  {journal} {Phys. Rev.}\ }\textbf {\bibinfo {volume}
  {100}},\ \bibinfo {pages} {564} (\bibinfo {year} {1955})}\BibitemShut
  {NoStop}%
\bibitem [{\citenamefont {Kanamori}(1959)}]{Kanamori1959}%
  \BibitemOpen
  \bibfield  {author} {\bibinfo {author} {\bibfnamefont {J.}~\bibnamefont
  {Kanamori}},\ }\href {\doibase 10.1016/0022-3697(59)90061-7} {\bibfield
  {journal} {\bibinfo  {journal} {J. Phys. Chem. Solids}\ }\textbf {\bibinfo
  {volume} {10}},\ \bibinfo {pages} {87} (\bibinfo {year} {1959})}\BibitemShut
  {NoStop}%
\bibitem [{\citenamefont {Blasse}(1965)}]{Blasse1965}%
  \BibitemOpen
  \bibfield  {author} {\bibinfo {author} {\bibfnamefont {G.}~\bibnamefont
  {Blasse}},\ }\href {\doibase 10.1016/0022-3697(65)90231-3} {\bibfield
  {journal} {\bibinfo  {journal} {J. Phys. Chem. Solids}\ }\textbf {\bibinfo
  {volume} {26}},\ \bibinfo {pages} {1969} (\bibinfo {year}
  {1965})}\BibitemShut {NoStop}%
\bibitem [{\citenamefont {Blonsky}\ \emph {et~al.}(2015)\citenamefont
  {Blonsky}, \citenamefont {Zhuang}, \citenamefont {Singh},\ and\ \citenamefont
  {Hennig}}]{Blonsky2020}%
  \BibitemOpen
  \bibfield  {author} {\bibinfo {author} {\bibfnamefont {M.~N.}\ \bibnamefont
  {Blonsky}}, \bibinfo {author} {\bibfnamefont {H.~L.}\ \bibnamefont {Zhuang}},
  \bibinfo {author} {\bibfnamefont {A.~K.}\ \bibnamefont {Singh}}, \ and\
  \bibinfo {author} {\bibfnamefont {R.~G.}\ \bibnamefont {Hennig}},\ }\href
  {\doibase 10.1021/acsnano.5b03394} {\bibfield  {journal} {\bibinfo  {journal}
  {ACS Nano}\ }\textbf {\bibinfo {volume} {9}},\ \bibinfo {pages} {9885}
  (\bibinfo {year} {2015})}\BibitemShut {NoStop}%
\bibitem [{\citenamefont {Duerloo}\ \emph {et~al.}(2012)\citenamefont
  {Duerloo}, \citenamefont {Ong},\ and\ \citenamefont {Reed}}]{Duerloo2012}%
  \BibitemOpen
  \bibfield  {author} {\bibinfo {author} {\bibfnamefont {K.-A.~N.}\
  \bibnamefont {Duerloo}}, \bibinfo {author} {\bibfnamefont {M.~T.}\
  \bibnamefont {Ong}}, \ and\ \bibinfo {author} {\bibfnamefont {E.~J.}\
  \bibnamefont {Reed}},\ }\href {\doibase 10.1021/jz3012436} {\bibfield
  {journal} {\bibinfo  {journal} {J. Phys. Chem. Lett.}\ }\textbf {\bibinfo
  {volume} {3}},\ \bibinfo {pages} {2871} (\bibinfo {year} {2012})}\BibitemShut
  {NoStop}%
\bibitem [{\citenamefont {{De Jong}}\ \emph {et~al.}(2015)\citenamefont {{De
  Jong}}, \citenamefont {Chen}, \citenamefont {Geerlings}, \citenamefont
  {Asta},\ and\ \citenamefont {Persson}}]{DeJong2015}%
  \BibitemOpen
  \bibfield  {author} {\bibinfo {author} {\bibfnamefont {M.}~\bibnamefont {{De
  Jong}}}, \bibinfo {author} {\bibfnamefont {W.}~\bibnamefont {Chen}}, \bibinfo
  {author} {\bibfnamefont {H.}~\bibnamefont {Geerlings}}, \bibinfo {author}
  {\bibfnamefont {M.}~\bibnamefont {Asta}}, \ and\ \bibinfo {author}
  {\bibfnamefont {K.~A.}\ \bibnamefont {Persson}},\ }\href {\doibase
  10.1038/sdata.2015.53} {\bibfield  {journal} {\bibinfo  {journal} {Sci. Data
  2015 21}\ }\textbf {\bibinfo {volume} {2}},\ \bibinfo {pages} {1} (\bibinfo
  {year} {2015})}\BibitemShut {NoStop}%
\bibitem [{\citenamefont {Toth}\ and\ \citenamefont {Lake}(2015)}]{Toth2015}%
  \BibitemOpen
  \bibfield  {author} {\bibinfo {author} {\bibfnamefont {S.}~\bibnamefont
  {Toth}}\ and\ \bibinfo {author} {\bibfnamefont {B.}~\bibnamefont {Lake}},\
  }\href {\doibase 10.1088/0953-8984/27/16/166002} {\bibfield  {journal}
  {\bibinfo  {journal} {J. Phys. Condens. Matter}\ }\textbf {\bibinfo {volume}
  {27}},\ \bibinfo {pages} {10} (\bibinfo {year} {2015})},\ \Eprint
  {http://arxiv.org/abs/1402.6069} {1402.6069} \BibitemShut {NoStop}%
\bibitem [{Sup({\natexlab{b}})}]{Suppcode}%
  \BibitemOpen
  \href@noop {} {} ({\natexlab{b}}),\ \bibinfo {note} {link to codes and
  outputs using Zenodo}\BibitemShut {NoStop}%
\bibitem [{\citenamefont {Mitra}\ \emph {et~al.}(2020)\citenamefont {Mitra},
  \citenamefont {Srivastava}, \citenamefont {Singha}, \citenamefont {Dutta},
  \citenamefont {Satpati}, \citenamefont {Karppinen}, \citenamefont {Ghosh},\
  and\ \citenamefont {Singha}}]{Mitra2020}%
  \BibitemOpen
  \bibfield  {author} {\bibinfo {author} {\bibfnamefont {S.}~\bibnamefont
  {Mitra}}, \bibinfo {author} {\bibfnamefont {D.}~\bibnamefont {Srivastava}},
  \bibinfo {author} {\bibfnamefont {S.~S.}\ \bibnamefont {Singha}}, \bibinfo
  {author} {\bibfnamefont {S.}~\bibnamefont {Dutta}}, \bibinfo {author}
  {\bibfnamefont {B.}~\bibnamefont {Satpati}}, \bibinfo {author} {\bibfnamefont
  {M.}~\bibnamefont {Karppinen}}, \bibinfo {author} {\bibfnamefont
  {A.}~\bibnamefont {Ghosh}}, \ and\ \bibinfo {author} {\bibfnamefont
  {A.}~\bibnamefont {Singha}},\ }\href {\doibase 10.1038/s41699-020-0138-y}
  {\bibfield  {journal} {\bibinfo  {journal} {npj 2D Mater. Appl.}\ }\textbf
  {\bibinfo {volume} {4}},\ \bibinfo {pages} {6} (\bibinfo {year}
  {2020})}\BibitemShut {NoStop}%
\bibitem [{\citenamefont {Wang}\ \emph {et~al.}(1993)\citenamefont {Wang},
  \citenamefont {Wu},\ and\ \citenamefont {Freeman}}]{Wang1993}%
  \BibitemOpen
  \bibfield  {author} {\bibinfo {author} {\bibfnamefont {D.~S.}\ \bibnamefont
  {Wang}}, \bibinfo {author} {\bibfnamefont {R.}~\bibnamefont {Wu}}, \ and\
  \bibinfo {author} {\bibfnamefont {A.~J.}\ \bibnamefont {Freeman}},\ }\href
  {\doibase 10.1103/PhysRevB.47.14932} {\bibfield  {journal} {\bibinfo
  {journal} {Phys. Rev. B}\ }\textbf {\bibinfo {volume} {47}},\ \bibinfo
  {pages} {14932} (\bibinfo {year} {1993})}\BibitemShut {NoStop}%
\bibitem [{\citenamefont {Kim}\ \emph {et~al.}(2020)\citenamefont {Kim},
  \citenamefont {Kim}, \citenamefont {Kim}, \citenamefont {Kang}, \citenamefont
  {Shin}, \citenamefont {Lee}, \citenamefont {Min},\ and\ \citenamefont
  {Park}}]{Kim2020a}%
  \BibitemOpen
  \bibfield  {author} {\bibinfo {author} {\bibfnamefont {J.}~\bibnamefont
  {Kim}}, \bibinfo {author} {\bibfnamefont {K.~W.}\ \bibnamefont {Kim}},
  \bibinfo {author} {\bibfnamefont {B.}~\bibnamefont {Kim}}, \bibinfo {author}
  {\bibfnamefont {C.~J.}\ \bibnamefont {Kang}}, \bibinfo {author}
  {\bibfnamefont {D.}~\bibnamefont {Shin}}, \bibinfo {author} {\bibfnamefont
  {S.~H.}\ \bibnamefont {Lee}}, \bibinfo {author} {\bibfnamefont {B.~C.}\
  \bibnamefont {Min}}, \ and\ \bibinfo {author} {\bibfnamefont
  {N.}~\bibnamefont {Park}},\ }\href {\doibase 10.1021/acs.nanolett.9b03815}
  {\bibfield  {journal} {\bibinfo  {journal} {Nano Lett.}\ }\textbf {\bibinfo
  {volume} {20}},\ \bibinfo {pages} {929} (\bibinfo {year} {2020})}\BibitemShut
  {NoStop}%
\bibitem [{\citenamefont {Steiner}\ \emph {et~al.}(2016)\citenamefont
  {Steiner}, \citenamefont {Khmelevskyi}, \citenamefont {Marsmann},\ and\
  \citenamefont {Kresse}}]{Steiner2016}%
  \BibitemOpen
  \bibfield  {author} {\bibinfo {author} {\bibfnamefont {S.}~\bibnamefont
  {Steiner}}, \bibinfo {author} {\bibfnamefont {S.}~\bibnamefont
  {Khmelevskyi}}, \bibinfo {author} {\bibfnamefont {M.}~\bibnamefont
  {Marsmann}}, \ and\ \bibinfo {author} {\bibfnamefont {G.}~\bibnamefont
  {Kresse}},\ }\href {\doibase 10.1103/PhysRevB.93.224425} {\bibfield
  {journal} {\bibinfo  {journal} {Phys. Rev. B}\ }\textbf {\bibinfo {volume}
  {93}},\ \bibinfo {pages} {224425} (\bibinfo {year} {2016})}\BibitemShut
  {NoStop}%
\bibitem [{\citenamefont {Perdew}\ \emph {et~al.}(1996)\citenamefont {Perdew},
  \citenamefont {Burke},\ and\ \citenamefont
  {Ernzerhof}}]{pbe_perdew_generalized_1996}%
  \BibitemOpen
  \bibfield  {author} {\bibinfo {author} {\bibfnamefont {J.~P.}\ \bibnamefont
  {Perdew}}, \bibinfo {author} {\bibfnamefont {K.}~\bibnamefont {Burke}}, \
  and\ \bibinfo {author} {\bibfnamefont {M.}~\bibnamefont {Ernzerhof}},\ }\href
  {\doibase 10.1103/PhysRevLett.77.3865} {\bibfield  {journal} {\bibinfo
  {journal} {Physical Review Letters}\ }\textbf {\bibinfo {volume} {77}},\
  \bibinfo {pages} {3865} (\bibinfo {year} {1996})}\BibitemShut {NoStop}%
\bibitem [{\citenamefont {Kresse}\ and\ \citenamefont
  {Furthm{\"{u}}ller}(1996{\natexlab{a}})}]{Kresse1996}%
  \BibitemOpen
  \bibfield  {author} {\bibinfo {author} {\bibfnamefont {G.}~\bibnamefont
  {Kresse}}\ and\ \bibinfo {author} {\bibfnamefont {J.}~\bibnamefont
  {Furthm{\"{u}}ller}},\ }\href
  {http://www.sciencedirect.com/science/article/pii/0927025696000080}
  {\bibfield  {journal} {\bibinfo  {journal} {Comput. Mater. Sci.}\ }\textbf
  {\bibinfo {volume} {6}},\ \bibinfo {pages} {15} (\bibinfo {year}
  {1996}{\natexlab{a}})}\BibitemShut {NoStop}%
\bibitem [{\citenamefont {Kresse}\ and\ \citenamefont
  {Furthm{\"{u}}ller}(1996{\natexlab{b}})}]{Kresse1996a}%
  \BibitemOpen
  \bibfield  {author} {\bibinfo {author} {\bibfnamefont {G.}~\bibnamefont
  {Kresse}}\ and\ \bibinfo {author} {\bibfnamefont {J.}~\bibnamefont
  {Furthm{\"{u}}ller}},\ }\href {http://www.ncbi.nlm.nih.gov/pubmed/9984901}
  {\bibfield  {journal} {\bibinfo  {journal} {Phys. Rev. B. Condens. Matter}\
  }\textbf {\bibinfo {volume} {54}},\ \bibinfo {pages} {11169} (\bibinfo {year}
  {1996}{\natexlab{b}})}\BibitemShut {NoStop}%
\bibitem [{\citenamefont {Kresse}\ and\ \citenamefont
  {Joubert}(1999)}]{Kresse1999}%
  \BibitemOpen
  \bibfield  {author} {\bibinfo {author} {\bibfnamefont {G.}~\bibnamefont
  {Kresse}}\ and\ \bibinfo {author} {\bibfnamefont {D.}~\bibnamefont
  {Joubert}},\ }\href {http://prb.aps.org/abstract/PRB/v59/i3/p1758{\_}1}
  {\bibfield  {journal} {\bibinfo  {journal} {Phys. Rev. B}\ }\textbf {\bibinfo
  {volume} {59}},\ \bibinfo {pages} {11} (\bibinfo {year} {1999})}\BibitemShut
  {NoStop}%
\bibitem [{\citenamefont {Neugebauer}\ and\ \citenamefont
  {Scheffler}(1992)}]{Neugebauer1992}%
  \BibitemOpen
  \bibfield  {author} {\bibinfo {author} {\bibfnamefont {J.}~\bibnamefont
  {Neugebauer}}\ and\ \bibinfo {author} {\bibfnamefont {M.}~\bibnamefont
  {Scheffler}},\ }\href {\doibase 10.1103/PhysRevB.46.16067} {\bibfield
  {journal} {\bibinfo  {journal} {Phys. Rev. B}\ }\textbf {\bibinfo {volume}
  {46}},\ \bibinfo {pages} {16067} (\bibinfo {year} {1992})}\BibitemShut
  {NoStop}%
\bibitem [{\citenamefont {Mathew}\ \emph {et~al.}(2017)\citenamefont {Mathew},
  \citenamefont {Montoya}, \citenamefont {Faghaninia}, \citenamefont
  {Dwarakanath}, \citenamefont {Aykol}, \citenamefont {Tang}, \citenamefont
  {heng Chu}, \citenamefont {Smidt}, \citenamefont {Bocklund}, \citenamefont
  {Horton}, \citenamefont {Dagdelen}, \citenamefont {Wood}, \citenamefont
  {Liu}, \citenamefont {Neaton}, \citenamefont {Ong}, \citenamefont {Persson},\
  and\ \citenamefont {Jain}}]{Mathew2017}%
  \BibitemOpen
  \bibfield  {author} {\bibinfo {author} {\bibfnamefont {K.}~\bibnamefont
  {Mathew}}, \bibinfo {author} {\bibfnamefont {J.~H.}\ \bibnamefont {Montoya}},
  \bibinfo {author} {\bibfnamefont {A.}~\bibnamefont {Faghaninia}}, \bibinfo
  {author} {\bibfnamefont {S.}~\bibnamefont {Dwarakanath}}, \bibinfo {author}
  {\bibfnamefont {M.}~\bibnamefont {Aykol}}, \bibinfo {author} {\bibfnamefont
  {H.}~\bibnamefont {Tang}}, \bibinfo {author} {\bibfnamefont {I.}~\bibnamefont
  {heng Chu}}, \bibinfo {author} {\bibfnamefont {T.}~\bibnamefont {Smidt}},
  \bibinfo {author} {\bibfnamefont {B.}~\bibnamefont {Bocklund}}, \bibinfo
  {author} {\bibfnamefont {M.}~\bibnamefont {Horton}}, \bibinfo {author}
  {\bibfnamefont {J.}~\bibnamefont {Dagdelen}}, \bibinfo {author}
  {\bibfnamefont {B.}~\bibnamefont {Wood}}, \bibinfo {author} {\bibfnamefont
  {Z.~K.}\ \bibnamefont {Liu}}, \bibinfo {author} {\bibfnamefont
  {J.}~\bibnamefont {Neaton}}, \bibinfo {author} {\bibfnamefont {S.~P.}\
  \bibnamefont {Ong}}, \bibinfo {author} {\bibfnamefont {K.}~\bibnamefont
  {Persson}}, \ and\ \bibinfo {author} {\bibfnamefont {A.}~\bibnamefont
  {Jain}},\ }\href {\doibase 10.1016/j.commatsci.2017.07.030} {\bibfield
  {journal} {\bibinfo  {journal} {Comput. Mater. Sci.}\ }\textbf {\bibinfo
  {volume} {139}},\ \bibinfo {pages} {140} (\bibinfo {year}
  {2017})}\BibitemShut {NoStop}%
\bibitem [{\citenamefont {Ong}\ \emph {et~al.}(2013)\citenamefont {Ong},
  \citenamefont {Richards}, \citenamefont {Jain}, \citenamefont {Hautier},
  \citenamefont {Kocher}, \citenamefont {Cholia}, \citenamefont {Gunter},
  \citenamefont {Chevrier}, \citenamefont {Persson},\ and\ \citenamefont
  {Ceder}}]{Ong2013}%
  \BibitemOpen
  \bibfield  {author} {\bibinfo {author} {\bibfnamefont {S.~P.}\ \bibnamefont
  {Ong}}, \bibinfo {author} {\bibfnamefont {W.~D.}\ \bibnamefont {Richards}},
  \bibinfo {author} {\bibfnamefont {A.}~\bibnamefont {Jain}}, \bibinfo {author}
  {\bibfnamefont {G.}~\bibnamefont {Hautier}}, \bibinfo {author} {\bibfnamefont
  {M.}~\bibnamefont {Kocher}}, \bibinfo {author} {\bibfnamefont
  {S.}~\bibnamefont {Cholia}}, \bibinfo {author} {\bibfnamefont
  {D.}~\bibnamefont {Gunter}}, \bibinfo {author} {\bibfnamefont {V.~L.}\
  \bibnamefont {Chevrier}}, \bibinfo {author} {\bibfnamefont {K.~A.}\
  \bibnamefont {Persson}}, \ and\ \bibinfo {author} {\bibfnamefont
  {G.}~\bibnamefont {Ceder}},\ }\href {\doibase
  10.1016/j.commatsci.2012.10.028} {\bibfield  {journal} {\bibinfo  {journal}
  {Comput. Mater. Sci.}\ }\textbf {\bibinfo {volume} {68}},\ \bibinfo {pages}
  {314} (\bibinfo {year} {2013})}\BibitemShut {NoStop}%
\bibitem [{\citenamefont {Giannozzi}\ \emph {et~al.}(2009)\citenamefont
  {Giannozzi}, \citenamefont {Baroni}, \citenamefont {Bonini}, \citenamefont
  {Calandra}, \citenamefont {Car}, \citenamefont {Cavazzoni}, \citenamefont
  {Ceresoli}, \citenamefont {Chiarotti}, \citenamefont {Cococcioni},
  \citenamefont {Dabo}, \citenamefont {Corso}, \citenamefont {de~Gironcoli},
  \citenamefont {Fabris}, \citenamefont {Fratesi}, \citenamefont {Gebauer},
  \citenamefont {Gerstmann}, \citenamefont {Gougoussis}, \citenamefont
  {Kokalj}, \citenamefont {Lazzeri}, \citenamefont {Martin-Samos},
  \citenamefont {Marzari}, \citenamefont {Mauri}, \citenamefont {Mazzarello},
  \citenamefont {Paolini}, \citenamefont {Pasquarello}, \citenamefont
  {Paulatto}, \citenamefont {Sbraccia}, \citenamefont {Scandolo}, \citenamefont
  {Sclauzero}, \citenamefont {Seitsonen}, \citenamefont {Smogunov},
  \citenamefont {Umari},\ and\ \citenamefont {Wentzcovitch}}]{Giannozzi2009}%
  \BibitemOpen
  \bibfield  {author} {\bibinfo {author} {\bibfnamefont {P.}~\bibnamefont
  {Giannozzi}}, \bibinfo {author} {\bibfnamefont {S.}~\bibnamefont {Baroni}},
  \bibinfo {author} {\bibfnamefont {N.}~\bibnamefont {Bonini}}, \bibinfo
  {author} {\bibfnamefont {M.}~\bibnamefont {Calandra}}, \bibinfo {author}
  {\bibfnamefont {R.}~\bibnamefont {Car}}, \bibinfo {author} {\bibfnamefont
  {C.}~\bibnamefont {Cavazzoni}}, \bibinfo {author} {\bibfnamefont
  {D.}~\bibnamefont {Ceresoli}}, \bibinfo {author} {\bibfnamefont {G.~L.}\
  \bibnamefont {Chiarotti}}, \bibinfo {author} {\bibfnamefont {M.}~\bibnamefont
  {Cococcioni}}, \bibinfo {author} {\bibfnamefont {I.}~\bibnamefont {Dabo}},
  \bibinfo {author} {\bibfnamefont {A.~D.}\ \bibnamefont {Corso}}, \bibinfo
  {author} {\bibfnamefont {S.}~\bibnamefont {de~Gironcoli}}, \bibinfo {author}
  {\bibfnamefont {S.}~\bibnamefont {Fabris}}, \bibinfo {author} {\bibfnamefont
  {G.}~\bibnamefont {Fratesi}}, \bibinfo {author} {\bibfnamefont
  {R.}~\bibnamefont {Gebauer}}, \bibinfo {author} {\bibfnamefont
  {U.}~\bibnamefont {Gerstmann}}, \bibinfo {author} {\bibfnamefont
  {C.}~\bibnamefont {Gougoussis}}, \bibinfo {author} {\bibfnamefont
  {A.}~\bibnamefont {Kokalj}}, \bibinfo {author} {\bibfnamefont
  {M.}~\bibnamefont {Lazzeri}}, \bibinfo {author} {\bibfnamefont
  {L.}~\bibnamefont {Martin-Samos}}, \bibinfo {author} {\bibfnamefont
  {N.}~\bibnamefont {Marzari}}, \bibinfo {author} {\bibfnamefont
  {F.}~\bibnamefont {Mauri}}, \bibinfo {author} {\bibfnamefont
  {R.}~\bibnamefont {Mazzarello}}, \bibinfo {author} {\bibfnamefont
  {S.}~\bibnamefont {Paolini}}, \bibinfo {author} {\bibfnamefont
  {A.}~\bibnamefont {Pasquarello}}, \bibinfo {author} {\bibfnamefont
  {L.}~\bibnamefont {Paulatto}}, \bibinfo {author} {\bibfnamefont
  {C.}~\bibnamefont {Sbraccia}}, \bibinfo {author} {\bibfnamefont
  {S.}~\bibnamefont {Scandolo}}, \bibinfo {author} {\bibfnamefont
  {G.}~\bibnamefont {Sclauzero}}, \bibinfo {author} {\bibfnamefont {A.~P.}\
  \bibnamefont {Seitsonen}}, \bibinfo {author} {\bibfnamefont {A.}~\bibnamefont
  {Smogunov}}, \bibinfo {author} {\bibfnamefont {P.}~\bibnamefont {Umari}}, \
  and\ \bibinfo {author} {\bibfnamefont {R.~M.}\ \bibnamefont {Wentzcovitch}},\
  }\href {\doibase 10.1088/0953-8984/21/39/395502} {\bibfield  {journal}
  {\bibinfo  {journal} {Journal of Physics: Condensed Matter}\ }\textbf
  {\bibinfo {volume} {21}},\ \bibinfo {pages} {395502} (\bibinfo {year}
  {2009})}\BibitemShut {NoStop}%
\bibitem [{\citenamefont {{Dal Corso}}(2014)}]{DalCorso2014}%
  \BibitemOpen
  \bibfield  {author} {\bibinfo {author} {\bibfnamefont {A.}~\bibnamefont {{Dal
  Corso}}},\ }\href {\doibase 10.1016/j.commatsci.2014.07.043} {\bibfield
  {journal} {\bibinfo  {journal} {Comput. Mater. Sci.}\ }\textbf {\bibinfo
  {volume} {95}},\ \bibinfo {pages} {337} (\bibinfo {year} {2014})}\BibitemShut
  {NoStop}%
\bibitem [{\citenamefont {Lejaeghere}\ \emph {et~al.}(2016)\citenamefont
  {Lejaeghere}, \citenamefont {Bihlmayer}, \citenamefont {Björkman},
  \citenamefont {Blaha}, \citenamefont {Blügel}, \citenamefont {Blum},
  \citenamefont {Caliste}, \citenamefont {Castelli}, \citenamefont {Clark},
  \citenamefont {Corso}, \citenamefont {de~Gironcoli}, \citenamefont {Deutsch},
  \citenamefont {Dewhurst}, \citenamefont {Marco}, \citenamefont {Draxl},
  \citenamefont {Dułak}, \citenamefont {Eriksson}, \citenamefont
  {Flores-Livas}, \citenamefont {Garrity}, \citenamefont {Genovese},
  \citenamefont {Giannozzi}, \citenamefont {Giantomassi}, \citenamefont
  {Goedecker}, \citenamefont {Gonze}, \citenamefont {Grånäs}, \citenamefont
  {Gross}, \citenamefont {Gulans}, \citenamefont {Gygi}, \citenamefont
  {Hamann}, \citenamefont {Hasnip}, \citenamefont {Holzwarth}, \citenamefont
  {Iuşan}, \citenamefont {Jochym}, \citenamefont {Jollet}, \citenamefont
  {Jones}, \citenamefont {Kresse}, \citenamefont {Koepernik}, \citenamefont
  {Küçükbenli}, \citenamefont {Kvashnin}, \citenamefont {Locht},
  \citenamefont {Lubeck}, \citenamefont {Marsman}, \citenamefont {Marzari},
  \citenamefont {Nitzsche}, \citenamefont {Nordström}, \citenamefont {Ozaki},
  \citenamefont {Paulatto}, \citenamefont {Pickard}, \citenamefont {Poelmans},
  \citenamefont {Probert}, \citenamefont {Refson}, \citenamefont {Richter},
  \citenamefont {Rignanese}, \citenamefont {Saha}, \citenamefont {Scheffler},
  \citenamefont {Schlipf}, \citenamefont {Schwarz}, \citenamefont {Sharma},
  \citenamefont {Tavazza}, \citenamefont {Thunström}, \citenamefont
  {Tkatchenko}, \citenamefont {Torrent}, \citenamefont {Vanderbilt},
  \citenamefont {van Setten}, \citenamefont {Speybroeck}, \citenamefont
  {Wills}, \citenamefont {Yates}, \citenamefont {Zhang},\ and\ \citenamefont
  {Cottenier}}]{Lejaeghere2016}%
  \BibitemOpen
  \bibfield  {author} {\bibinfo {author} {\bibfnamefont {K.}~\bibnamefont
  {Lejaeghere}}, \bibinfo {author} {\bibfnamefont {G.}~\bibnamefont
  {Bihlmayer}}, \bibinfo {author} {\bibfnamefont {T.}~\bibnamefont
  {Björkman}}, \bibinfo {author} {\bibfnamefont {P.}~\bibnamefont {Blaha}},
  \bibinfo {author} {\bibfnamefont {S.}~\bibnamefont {Blügel}}, \bibinfo
  {author} {\bibfnamefont {V.}~\bibnamefont {Blum}}, \bibinfo {author}
  {\bibfnamefont {D.}~\bibnamefont {Caliste}}, \bibinfo {author} {\bibfnamefont
  {I.~E.}\ \bibnamefont {Castelli}}, \bibinfo {author} {\bibfnamefont {S.~J.}\
  \bibnamefont {Clark}}, \bibinfo {author} {\bibfnamefont {A.~D.}\ \bibnamefont
  {Corso}}, \bibinfo {author} {\bibfnamefont {S.}~\bibnamefont {de~Gironcoli}},
  \bibinfo {author} {\bibfnamefont {T.}~\bibnamefont {Deutsch}}, \bibinfo
  {author} {\bibfnamefont {J.~K.}\ \bibnamefont {Dewhurst}}, \bibinfo {author}
  {\bibfnamefont {I.~D.}\ \bibnamefont {Marco}}, \bibinfo {author}
  {\bibfnamefont {C.}~\bibnamefont {Draxl}}, \bibinfo {author} {\bibfnamefont
  {M.}~\bibnamefont {Dułak}}, \bibinfo {author} {\bibfnamefont
  {O.}~\bibnamefont {Eriksson}}, \bibinfo {author} {\bibfnamefont {J.~A.}\
  \bibnamefont {Flores-Livas}}, \bibinfo {author} {\bibfnamefont {K.~F.}\
  \bibnamefont {Garrity}}, \bibinfo {author} {\bibfnamefont {L.}~\bibnamefont
  {Genovese}}, \bibinfo {author} {\bibfnamefont {P.}~\bibnamefont {Giannozzi}},
  \bibinfo {author} {\bibfnamefont {M.}~\bibnamefont {Giantomassi}}, \bibinfo
  {author} {\bibfnamefont {S.}~\bibnamefont {Goedecker}}, \bibinfo {author}
  {\bibfnamefont {X.}~\bibnamefont {Gonze}}, \bibinfo {author} {\bibfnamefont
  {O.}~\bibnamefont {Grånäs}}, \bibinfo {author} {\bibfnamefont {E.~K.~U.}\
  \bibnamefont {Gross}}, \bibinfo {author} {\bibfnamefont {A.}~\bibnamefont
  {Gulans}}, \bibinfo {author} {\bibfnamefont {F.}~\bibnamefont {Gygi}},
  \bibinfo {author} {\bibfnamefont {D.~R.}\ \bibnamefont {Hamann}}, \bibinfo
  {author} {\bibfnamefont {P.~J.}\ \bibnamefont {Hasnip}}, \bibinfo {author}
  {\bibfnamefont {N.~A.~W.}\ \bibnamefont {Holzwarth}}, \bibinfo {author}
  {\bibfnamefont {D.}~\bibnamefont {Iuşan}}, \bibinfo {author} {\bibfnamefont
  {D.~B.}\ \bibnamefont {Jochym}}, \bibinfo {author} {\bibfnamefont
  {F.}~\bibnamefont {Jollet}}, \bibinfo {author} {\bibfnamefont
  {D.}~\bibnamefont {Jones}}, \bibinfo {author} {\bibfnamefont
  {G.}~\bibnamefont {Kresse}}, \bibinfo {author} {\bibfnamefont
  {K.}~\bibnamefont {Koepernik}}, \bibinfo {author} {\bibfnamefont
  {E.}~\bibnamefont {Küçükbenli}}, \bibinfo {author} {\bibfnamefont {Y.~O.}\
  \bibnamefont {Kvashnin}}, \bibinfo {author} {\bibfnamefont {I.~L.~M.}\
  \bibnamefont {Locht}}, \bibinfo {author} {\bibfnamefont {S.}~\bibnamefont
  {Lubeck}}, \bibinfo {author} {\bibfnamefont {M.}~\bibnamefont {Marsman}},
  \bibinfo {author} {\bibfnamefont {N.}~\bibnamefont {Marzari}}, \bibinfo
  {author} {\bibfnamefont {U.}~\bibnamefont {Nitzsche}}, \bibinfo {author}
  {\bibfnamefont {L.}~\bibnamefont {Nordström}}, \bibinfo {author}
  {\bibfnamefont {T.}~\bibnamefont {Ozaki}}, \bibinfo {author} {\bibfnamefont
  {L.}~\bibnamefont {Paulatto}}, \bibinfo {author} {\bibfnamefont {C.~J.}\
  \bibnamefont {Pickard}}, \bibinfo {author} {\bibfnamefont {W.}~\bibnamefont
  {Poelmans}}, \bibinfo {author} {\bibfnamefont {M.~I.~J.}\ \bibnamefont
  {Probert}}, \bibinfo {author} {\bibfnamefont {K.}~\bibnamefont {Refson}},
  \bibinfo {author} {\bibfnamefont {M.}~\bibnamefont {Richter}}, \bibinfo
  {author} {\bibfnamefont {G.-M.}\ \bibnamefont {Rignanese}}, \bibinfo {author}
  {\bibfnamefont {S.}~\bibnamefont {Saha}}, \bibinfo {author} {\bibfnamefont
  {M.}~\bibnamefont {Scheffler}}, \bibinfo {author} {\bibfnamefont
  {M.}~\bibnamefont {Schlipf}}, \bibinfo {author} {\bibfnamefont
  {K.}~\bibnamefont {Schwarz}}, \bibinfo {author} {\bibfnamefont
  {S.}~\bibnamefont {Sharma}}, \bibinfo {author} {\bibfnamefont
  {F.}~\bibnamefont {Tavazza}}, \bibinfo {author} {\bibfnamefont
  {P.}~\bibnamefont {Thunström}}, \bibinfo {author} {\bibfnamefont
  {A.}~\bibnamefont {Tkatchenko}}, \bibinfo {author} {\bibfnamefont
  {M.}~\bibnamefont {Torrent}}, \bibinfo {author} {\bibfnamefont
  {D.}~\bibnamefont {Vanderbilt}}, \bibinfo {author} {\bibfnamefont {M.~J.}\
  \bibnamefont {van Setten}}, \bibinfo {author} {\bibfnamefont {V.~V.}\
  \bibnamefont {Speybroeck}}, \bibinfo {author} {\bibfnamefont {J.~M.}\
  \bibnamefont {Wills}}, \bibinfo {author} {\bibfnamefont {J.~R.}\ \bibnamefont
  {Yates}}, \bibinfo {author} {\bibfnamefont {G.-X.}\ \bibnamefont {Zhang}}, \
  and\ \bibinfo {author} {\bibfnamefont {S.}~\bibnamefont {Cottenier}},\ }\href
  {\doibase 10.1126/science.aad3000} {\bibfield  {journal} {\bibinfo  {journal}
  {Science}\ }\textbf {\bibinfo {volume} {351}},\ \bibinfo {pages} {aad3000}
  (\bibinfo {year} {2016})}\BibitemShut {NoStop}%
\bibitem [{\citenamefont {Baroni}\ \emph {et~al.}(1987)\citenamefont {Baroni},
  \citenamefont {Giannozzi},\ and\ \citenamefont {Testa}}]{Baroni1987}%
  \BibitemOpen
  \bibfield  {author} {\bibinfo {author} {\bibfnamefont {S.}~\bibnamefont
  {Baroni}}, \bibinfo {author} {\bibfnamefont {P.}~\bibnamefont {Giannozzi}}, \
  and\ \bibinfo {author} {\bibfnamefont {A.}~\bibnamefont {Testa}},\ }\href
  {\doibase 10.1103/PhysRevLett.58.1861} {\bibfield  {journal} {\bibinfo
  {journal} {Phys. Rev. Lett.}\ }\textbf {\bibinfo {volume} {58}},\ \bibinfo
  {pages} {1861} (\bibinfo {year} {1987})}\BibitemShut {NoStop}%
\bibitem [{\citenamefont {Baroni}\ \emph {et~al.}(2001)\citenamefont {Baroni},
  \citenamefont {de~Gironcoli}, \citenamefont {Dal~Corso},\ and\ \citenamefont
  {Giannozzi}}]{Baroni2001}%
  \BibitemOpen
  \bibfield  {author} {\bibinfo {author} {\bibfnamefont {S.}~\bibnamefont
  {Baroni}}, \bibinfo {author} {\bibfnamefont {S.}~\bibnamefont
  {de~Gironcoli}}, \bibinfo {author} {\bibfnamefont {A.}~\bibnamefont
  {Dal~Corso}}, \ and\ \bibinfo {author} {\bibfnamefont {P.}~\bibnamefont
  {Giannozzi}},\ }\href {\doibase 10.1103/RevModPhys.73.515} {\bibfield
  {journal} {\bibinfo  {journal} {Rev. Mod. Phys.}\ }\textbf {\bibinfo {volume}
  {73}},\ \bibinfo {pages} {515} (\bibinfo {year} {2001})}\BibitemShut
  {NoStop}%
\end{thebibliography}%
\end{document}